\documentclass[lettersize,journal]{IEEEtran}
\usepackage{amsmath,amsfonts}
\usepackage{algorithmic}
\usepackage{algorithm}
\usepackage{array}
\usepackage[caption=false,font=normalsize,labelfont=sf,textfont=sf]{subfig}
\usepackage{textcomp}
\usepackage{stfloats}
\usepackage{url}
\usepackage{verbatim}
\usepackage{graphicx}
\usepackage{cite}
\usepackage[hidelinks,breaklinks]{hyperref}
\begin{document}

\title{AnatomyCarve: A VR occlusion management technique for medical images based on segment-aware clipping}

\author{Andrey~Titov,~Tina~N.~H.~Nantenaina,~Marta~Kersten-Oertel~and~Simon~Drouin
        % <-this % stops a space
\thanks{Andrey Titov, Tina N. H. Nantenaina and Simon Drouin are with Software and Information Technology Engineering department of École de technologie supérieure, Montreal, Canada.}
\thanks{Andrey Titov and Marta Kersten-Oertel are with the Gina Cody School of Computer Science Engineering, Concordia University, Montreal,
Canada.}% <-this % stops a space
%\thanks{Manuscript received April 19, 2021; revised August 16, 2021.}
}



\maketitle

\begin{abstract}
Visualizing 3D medical images is challenging due to self-occlusion, where anatomical structures of interest can be obscured by surrounding tissues. Existing methods, such as slicing and interactive clipping, are limited in their ability to fully represent internal anatomy in context. In contrast, hand-drawn medical illustrations in anatomy books manage occlusion effectively by selectively removing portions based on tissue type, revealing 3D structures while preserving context. This paper introduces AnatomyCarve, a novel technique developed for a VR environment that creates high-quality illustrations similar to those in anatomy books, while remaining fast and interactive. AnatomyCarve allows users to clip selected segments from 3D medical volumes, preserving spatial relations and contextual information. This approach enhances visualization by combining advanced rendering techniques with natural user interactions in VR. Usability of AnatomyCarve was assessed through a study with non-experts, while surgical planning effectiveness was evaluated with practicing neurosurgeons and residents. The results show that AnatomyCarve enables customized anatomical visualizations, with high user satisfaction, suggesting its potential for educational and clinical applications.
\end{abstract}


\section{Introduction}
\IEEEPARstart{V}{isualizing} three-dimensional medical images, such as those produced by MRI or CT scanners, poses significant challenges in effectively depicting internal anatomy primarily due to self-occlusion. In the context of anatomical images, self-occlusion occurs when certain anatomical structures obscure others within the body. For example, outer structures (e.g. skin) may block the view of internal ones (e.g. organs), making it difficult to discern detailed internal anatomy in the greater context of the human body.

%\begin{figure*}[!t]
%\centering
%\includegraphics[width=\linewidth]{figures/teaser2down2.png}
%\caption{Comparison between anatomical book illustrations on the far-left and and middle-right panels, and their recreation with our tool, AnatomyCarve, on the middle left and far-right panels correspondingly.}
%\label{fig:main}
%\end{figure*}

\begin{figure*}[!t]
\centering
\subfloat[]{\includegraphics[width=3.5in]{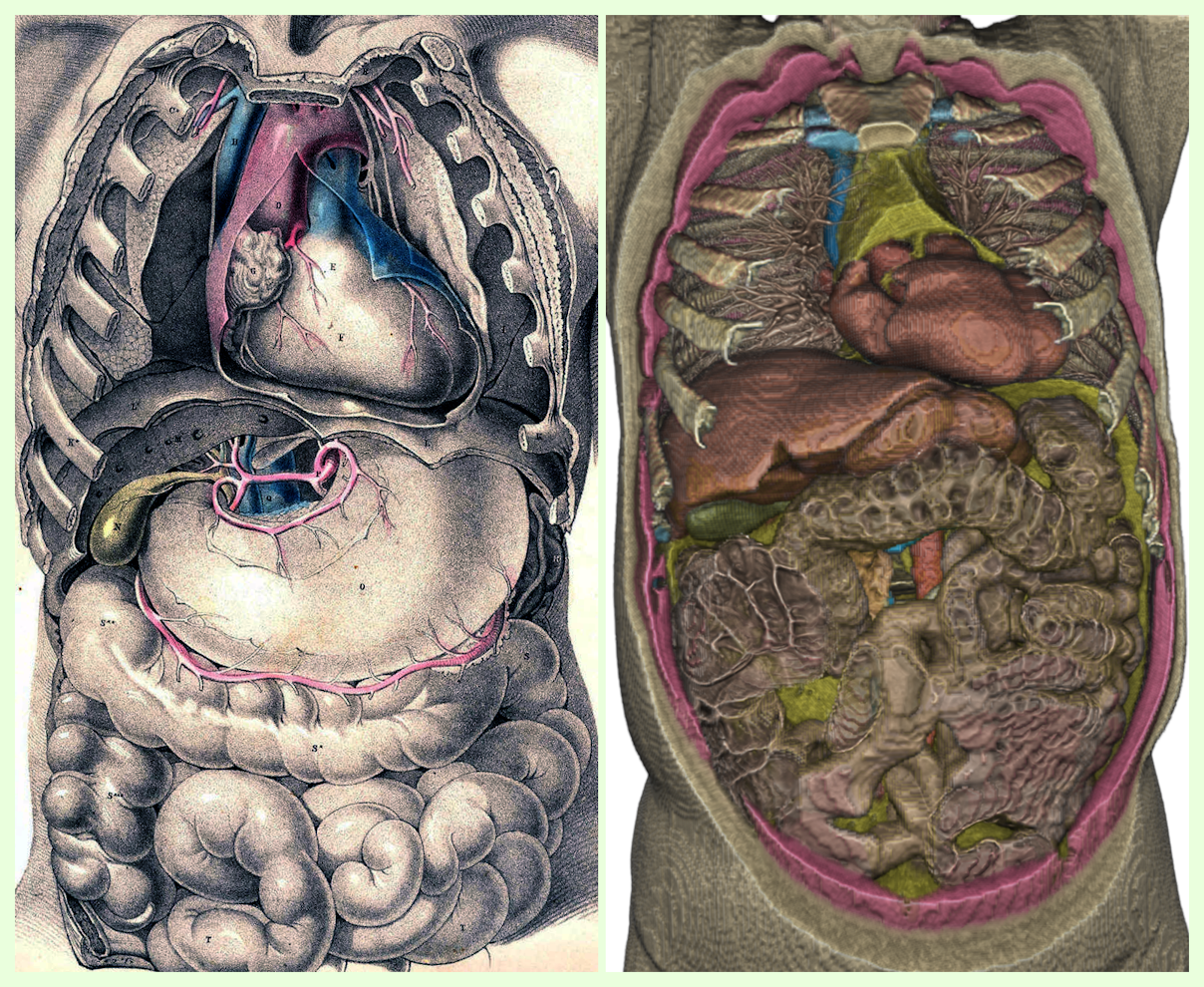}%
\label{fig:main_a}}
\hfil
\subfloat[]{\includegraphics[width=3.5in]{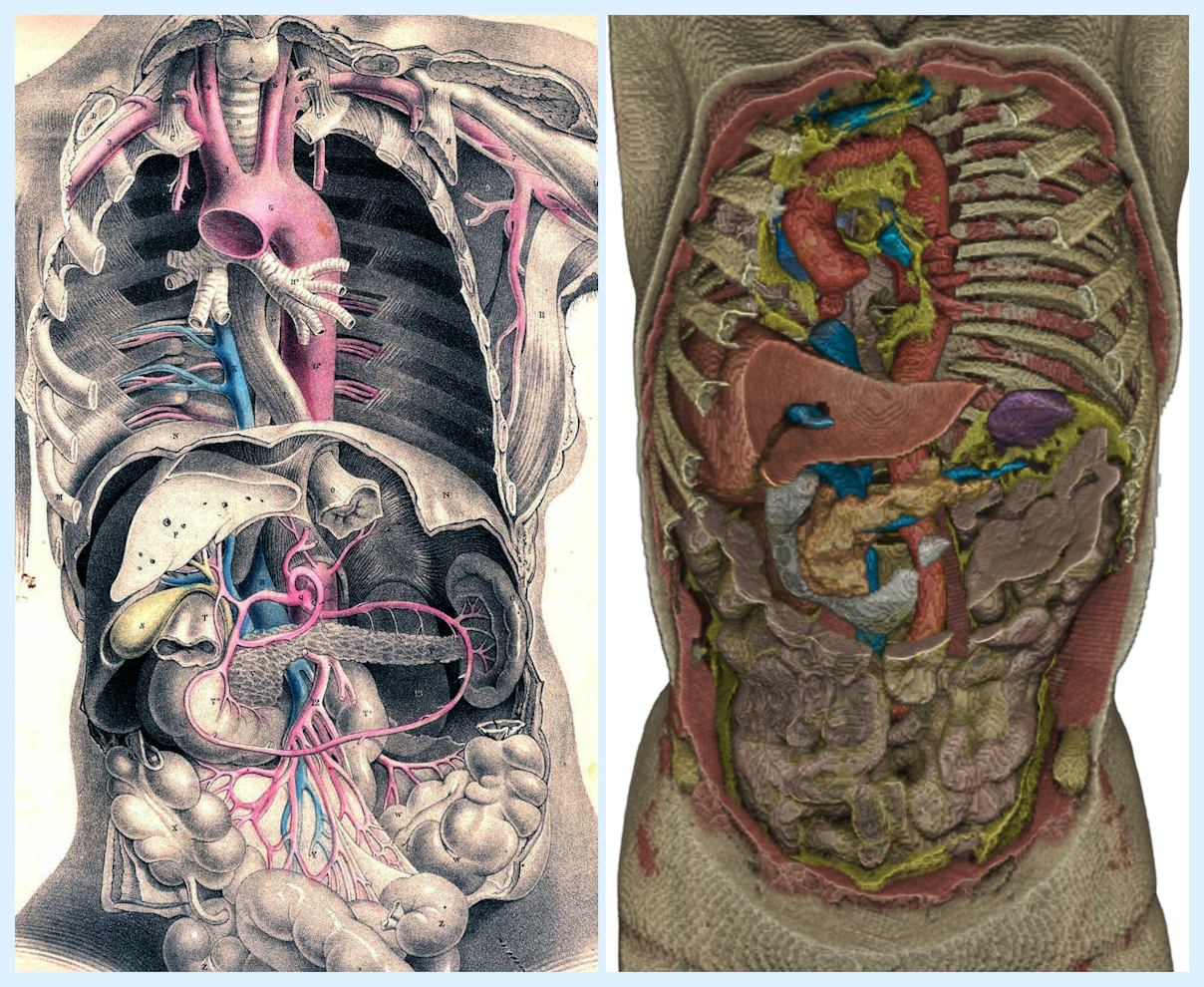}%
\label{fig:main_b}}
\caption{Comparison between anatomical book illustrations on the left panels in (a) and (b), and their recreation with our tool, AnatomyCarve, on the right panels in (a) and (b) correspondingly. Illustrations on the left panels of (a) and (b) adapted from J.~Maclise, \textit{Surgical Anatomy} (1859), public domain \cite{maclise_surgical_1859}.}
\label{fig:main}
\end{figure*}

To address this problem, several solutions have been proposed, including axis-aligned slicing (AAS) \cite{joshi_novel_2008}, arbitrary slicing (AS) \cite{laha_volume_2013}, cut-away views \cite{viola_importance-driven_2004}, focus and context techniques \cite{wang_magic_2005}, interactive clipping \cite{joshi_novel_2008,laha_volume_2013}, and deformations \cite{mcguffin_using_2003}. AAS and AS work by displaying 2D slices that were sampled from the three-dimensional volume. Cut-away views and clipping techniques expose hidden structures by selectively removing occluding elements, while deformations convey how contextual layers (such as skin) can be moved aside to reveal internal anatomy.

The goal of these techniques is to enable interactive visualization that allows users to view different anatomical layers or structures within the broader context of the body. For example, Joshi et al. \cite{joshi_novel_2008} proposed a 3D method that used a clipping shape to hide a subset of the volume's voxels. However, their approach only reveals the anatomy that lies on the surface of the clipping mesh and does not facilitate a deeper understanding of the internal 3D structures. McGuffin et al. \cite{mcguffin_using_2003} introduced interactive deformation techniques that can be applied to segmented volumetric medical images. A key limitation of their approach is that, due to the movement of the different components, the 3D structures are no longer at their original position, making it more difficult to perceive and understand the spatial relations between the anatomical structures. 

In this paper, we present AnatomyCarve, a method that addresses key limitations of existing clipping and deformation techniques in the visualization of 3D medical images. Our approach was inspired by medical illustrations commonly found in anatomical textbooks (such as Gray's Anatomy \cite{standring_grays_2015}), which use segment-based clipping or cut-away views to reveal internal structures by selectively removing outer layers (see Fig. \ref{fig:main}). These illustrations provide a clear depiction of the spatial relationships between skin, organs, bones, and vessels, offering a comprehensive understanding of complex anatomy. AnatomyCarve brings this illustrative power into an interactive 3D environment by enabling real-time, segment-specific clipping while preserving the spatial context of anatomical structures. To our knowledge, no existing method supports per-organ or per-segment interactive clipping that doesn't rely on the focus and context paradigm, but rather employs the idea of "stacking", similar to the approach used in AnatomyCarve, where multiple clipping shapes are inserted in the volume, each carving a different segments. Additionally, we have introduced a new metric, called Mean Absolute Error (MAE) of the first encountered segment, that measures the similarity between two visualizations of the same image with same point of view, but with different elements being clipped.

AnatomyCarve was  developed specifically for VR to leverage its natural interaction and 3D visualization capabilities. VR headsets with controllers offering 6 degrees of freedom (6DoF) enable users to perform precise clipping operations more intuitively than traditional 2D input methods like a computer mouse \cite{jerald_vr_2015}. In VR, users can move the clipping shape freely in 3D space, offering greater flexibility. Moreover, VR provides strong depth cues, such as stereoscopy and motion parallax \cite{vienne_depth_2020}, which further aid in positioning clipping shapes and ensuring accurate segmentations during the creation of medical views. This combination of interactivity and advanced visualization makes AnatomyCarve a powerful tool for both educational and clinical applications.

We believe that AnatomyCarve can be a valuable tool for creating high-quality medical illustrations and teaching materials, helping to clarify complex anatomical relationships. In addition, within a virtual reality (VR) environment, AnatomyCarve serves as an immersive learning tool for studying human anatomy interactively, engaging users in a hands-on exploration of anatomical structures. This immersive approach enhances the learning experience by allowing users to intuitively navigate and manipulate 3D anatomical models. Furthermore, it holds significant potential for enhancing surgical planning by offering a more detailed understanding of patient-specific anatomy, ultimately improving the precision of surgical procedures. We will demonstrate the usefulness of AnatomyCarve for these tasks in this paper.

\section{Background and Related Work}

\subsection{Anatomical Illustration}

Anatomical illustration plays a crucial role in medical education and communication, serving not only as a pedagogical tool to convey complex spatial relationships and physiological structures, but also as an artistic medium that bridges science and visual storytelling \cite{weiss_leveraging_2023}. From an educational standpoint, illustrations can simplify intricate anatomical concepts, enhance memory retention, and support learners at various stages of training \cite{papa_teaching_2021}. Artistically, anatomical renderings have long provided a unique way to interpret and represent the human body, contributing to both scientific understanding and cultural heritage \cite{calkins_human_1999,naicker_medical_2023}.

Such illustrations are often designed with specific, task-dependent goals in mind \cite[p. 419]{preim_visualization_2007}. These illustrations focus on visualizing certain subsets of organs or tissues relevant to the task at hand, while less important elements are either partially or fully clipped to avoid occlusion.  Therefore, artists creating these illustrations carefully consider which anatomical elements should be preserved, which can be cut or partially removed to maintain spatial relationships, and which can be entirely excluded for the current focus \cite[p. 419]{preim_visualization_2007}. For example, in the illustrations presented in the left panels of Fig. \ref{fig:main_a} and  Fig. \ref{fig:main_b}, less important structures such as the ribs are partially retained to preserve the spatial relationships between the anatomical components, while outer tissues such as muscles or the skin are entirely removed to avoid obstructing the organs of the digestive system organs. In some cases, an organ may be cut to reveal its internal structure, as seen with the liver in the left panel of Fig. \ref{fig:main_b}. Additionally, deformations, such as those that can be seen in the left panel of Fig. \ref{fig:main_b}, may be used to reposition structures—such as moving the bowel outside the body—to reveal underlying organs. The  goal of these illustrations varies based on the target anatomical system, as evidenced by the left panel of Fig. \ref{fig:main_a}, which focuses on the digestive system, and the left panel of Fig. \ref{fig:main_b}, which focuses on the cardiovascular system. AnatomyCarve enables the creation of similar task-dependent visualizations, as can be seen in the right panels of Fig. \ref{fig:main_a} and \ref{fig:main_b}.

\subsection{Occlusion Management Techniques}
A number of solutions have been proposed for solving the self-occlusion problem, commonly referred to as occlusion management techniques \cite{elmqvist_taxonomy_2008} (see reference for detailed review). Among the earliest and most basic methods are slicing and clipping, which aim to reveal internal components of complex volumes that would otherwise be hidden when the entire volume is rendered simultaneously.

Slicing is typically implemented by sampling one or more planar slices from a 3D volume and then displaying them in 2D \cite{joshi_novel_2008}. The simplest form of slicing is axis-aligned slicing (AAS), where the slicing planes are aligned perpendicular to the three default axes in 3D, with each slice sampling the volume. A variation of this method, known as arbitrary slicing, allows for arbitrary 2D slices through the volume \cite{laha_volume_2013}. The drawback of slicing methods is that they make it difficult to fully comprehend the 3D structures of anatomical regions \cite[p. 155]{preim_visualization_2007}, particularly for elongated structures such as electrodes \cite{joshi_novel_2008} or blood vessels.

Clipping, in contrast, works by hiding parts of the dataset in order to visualize a smaller subset of a volume \cite[p. 291]{preim_visualization_2007}. This approach makes the data easier to understand, as it reduces the number of voxels that need to be visualized simultaneously. Additionally, by removing external layers, clipping can reveal internal structures that would otherwise remain obscured\cite{joshi_novel_2008,drouin_prism_2018}. One of the simplest forms of clipping is axis-aligned cropping, but it has the drawback that the mapping between the 2D slices and an arbitrary oriented volume may require significant mental effort. Furthermore, it may crop away too much of the surrounding region, making it harder to maintain spatial context \cite{joshi_novel_2008}.

\begin{figure}[ht]
\centering{\includegraphics[width=3.5in]{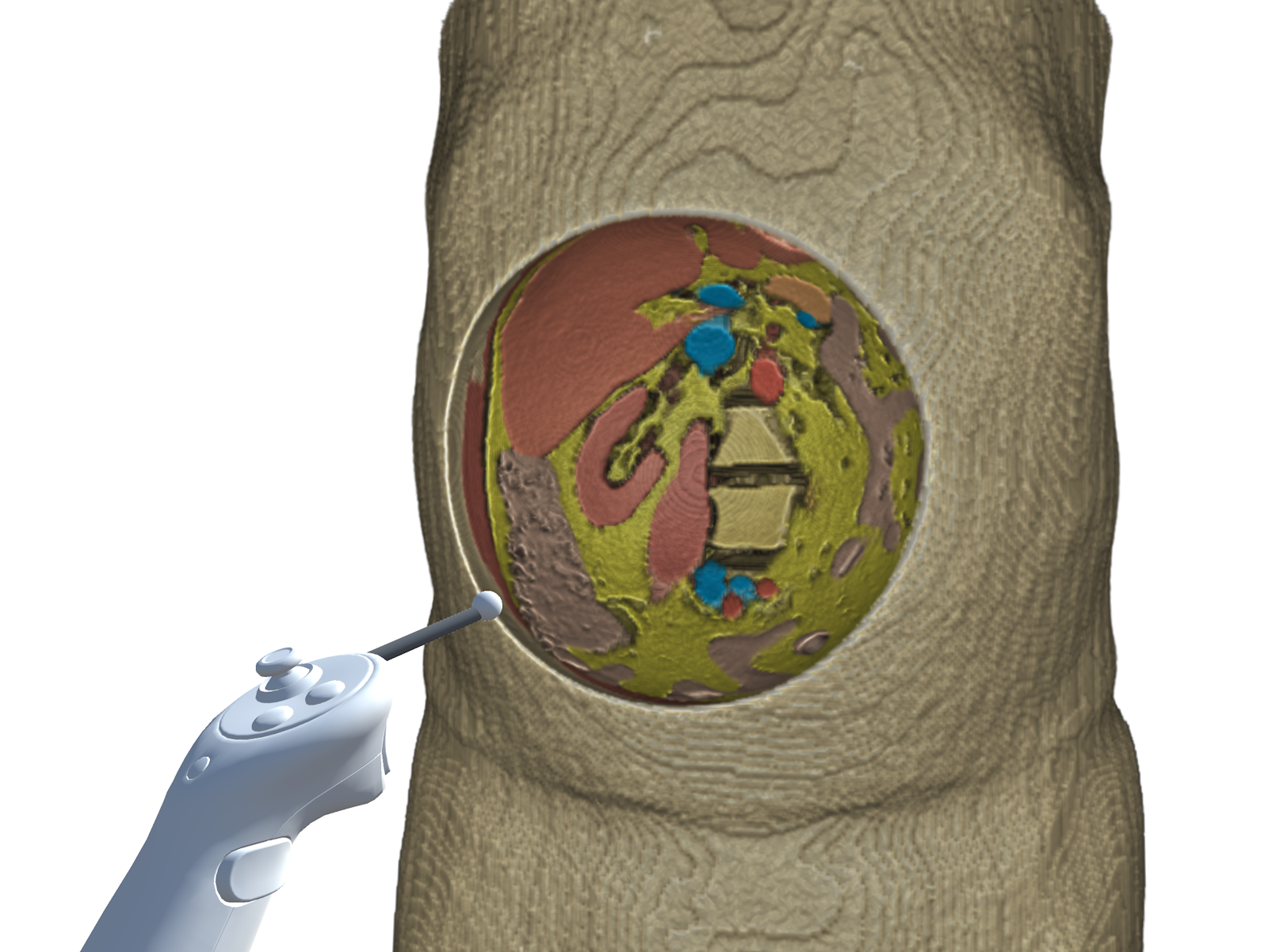}}
\caption{A simulation of the Joshi et al. \cite{joshi_novel_2008} clipping algorithm within AnatomyCarve. Note that the clipping doesn't discriminate between the segments and clips all the voxels within the spherical region.}
\label{fig:joshi-alike}
\end{figure}

Extending upon the idea of planar clipping, Ajani et al. \cite{ajani_volumetric_2018} introduced the idea of volumetric clipping surfaces that work on segmented data. The main idea of this technique is that a 2D spline surface area is inserted in the volume, clipping voxels on one side. The technique employs the focus and context paradigm, where one of the selected segments is kept as the focus, while the surrounding area is smoothly clipped around. Thus, like an invisible veil, the spline surface area wraps the focus segment, and then smoothly recedes into the surrounding organs and tissues. The technique requires no user input other than selecting the focus segment.

Joshi et al. \cite{joshi_novel_2008} explored the use of irregular cropping of anatomical volumes in the context of neurosurgery. They developed a neurosurgical visualization system in which the user could slice parts of the volume using three-dimensional convex shapes to reveal the internal 
parts of the brain. When such a shape was inserted in the brain, the voxels inside the shape were excluded from rendering during the ray casting stage. The advantage that irregular cropping has over axis-aligned slicing is that it preserves more context when visualizing the data by allowing users to crop only a small subset of the visible data. However, it does not enable the visualization of the 3D shape of the internal anatomical structures, instead, it only displays the anatomy on the surface of the clipping shape, as illustrated in Fig. \ref{fig:joshi-alike}.

McGuffin et al. \cite{mcguffin_using_2003} proposed a solution to the occlusion management problem by providing the viewer with a set of deformation tools that could be used to enhance data understanding. They developed a medical visualization system that allows users to view and manipulate a volumetric dataset. The system's interface allows the user to select a region within the volume, adjust the transfer function, and deform or spatially transform the selected region. The system offered 5 tools to manipulate and transform the data, which could be applied either to the entire volume or specific segments. Each manipulation was animated to ensure a smooth transition between different states. The advantage of these deformation-based techniques lies in their ability to reveal hidden interior parts of the dataset, while maintaining the relative positioning of the anatomical elements. The system used the computer mouse and keyboard as input devices, with various widgets to facilitate interaction. 

Díaz et al. \cite{diaz_adaptive_2012} introduced an algorithm based on interactive clipping that allows interactively creating cross-sections of medical images using a segment-based expansion. The algorithm works best with segmented data, although in the case that the segmentation is not provided, on-the-fly region growing is employed. The core idea of the interaction model used in this algorithm is that the user inserts a clipping plane in a medical image, which hides all voxels on one side of it one one side. Then, using a mouse, the user clicks on one of the remaining visible pixels, which selects the corresponding segment. After that, by dragging the mouse, the user can "expand" this segment in the clipped region, making its voxels reappear again. The technique was validated with an informal user study, where participants had to recreate test images that were presented to them using the system. It was found that the system has a short learning process, and was quick to use, as reproducing the different images took anywhere between a couple of seconds up to a few minutes for the participants.

Laha and Bowman \cite{laha_volume_2013} designed a gesture-based interaction technique called Volume Cracker, which built upon the deformation methods introduced by McGuffin et al. \cite{mcguffin_using_2003}. Similar to McGuffin et al., the goal of Volume Cracker was to enhance the understanding of three-dimensional projected data through deformations. However, instead of using a traditional computer mouse, Laha and Bowman \cite{laha_volume_2013} employed a bimanual, asymmetrical input technique using tracked gloves. In this approach, a splitting plane was positioned at the midpoint between the two hands, and a grabbing and separating gesture caused the volume to split into two pieces. Later, Laha et al. \cite{laha_bare-hand_2016} proposed an alternative version of Volume Cracker, replacing the tracked glove and controllers with a bare-hand interaction technique using a Leap Motion controller. They argued that, although Volume Cracker showed promise, the handheld controllers were impractical and limited the technique's accessibility.

Lastly, He et al. \cite{he_medical_2017} introduced a medical data visualization technique for exploring pre-segmented brain parts organized in a hierarchical structure. The system allowed users to create exploded views of anatomical regions, which facilitated the visualization of densely packed structures, such as those inside the brain. This technique was implemented in a VR environment and used handheld controllers. The controllers' positions generated points, which were used to compute a Hermite spline curve. These splines defined the trajectory for moving the segmented parts of the data. The trackpad could be used to reparametrize the curve, determining which segments would be presented in an exploded view and which would remain intact. The splines also tracked the orientation of the volume, optimizing the movement trajectory to minimize rotation of the segments. The deformations applied to the dataset were linear and were inspired by the techniques described by McGuffin et al. \cite{mcguffin_using_2003}.

Building on these previous works, we have developed a novel interactive occlusion management method that seamlessly integrates the advantages of both interactive clipping and deformation, while overcoming the limitations inherent in each approach.

\section{Methods}
In this section, we begin by outlining the core functionality of AnatomyCarve, followed by an overview of the interaction methods within the VR environment, and then we explain the rendering process.

\subsection{Core Concept}

\begin{figure}[ht]
\centering{\includegraphics[width=3.5in]{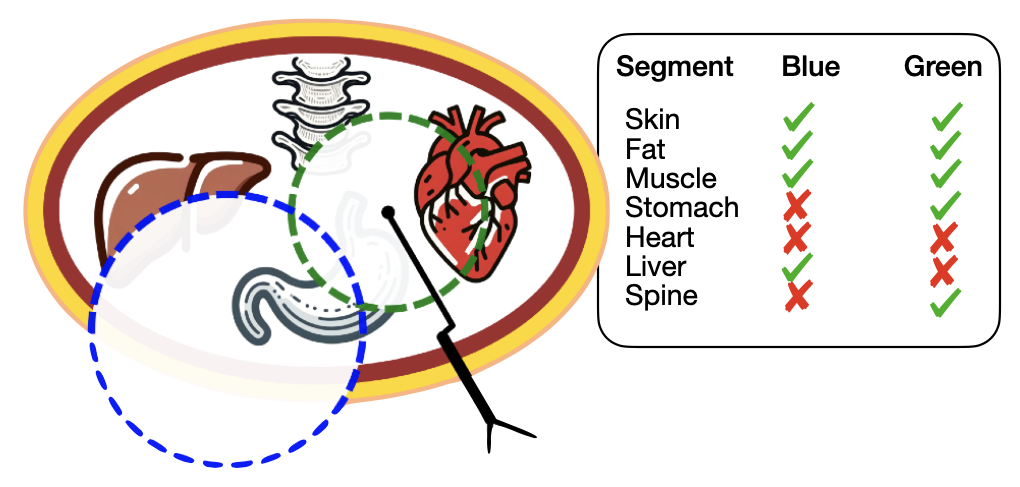}}
\caption{A carving configuration with two clipping spheres. The sphere represented by the blue color clips the skin, fat, muscle, and liver. At the same time, the sphere represented by the green color also clips the skin fat and muscle, but instead of the liver it clips the stomach and the spine. The table shows the corresponding clipping mask, where check marks indicate clipped segments and crosses indicate unaltered ones.}
\label{fig:sketch}
\end{figure}

Traditional medical illustrations address the self-occlusion problem effectively by selectively omitting or transparently depicting overlapping anatomical structures to maintain clarity and comprehension. Similarly, our method employs a comparable philosophy, allowing the user to decide which organs to clip and which ones to leave untouched based on their specific use case. Fig. \ref{fig:sketch} illustrates an anatomical carving configuration, where the user applied two distinct cuts to the volume, represented by circles (blue and green) that correspond to three-dimensional spheres in AnatomyCarve. Spherical shapes were chosen for their simplicity, as they can easily be combined to form more complex cuts. In the example shown in Fig. \ref{fig:sketch}, the user first positioned a spherical cutting shape (blue dotted circle) to remove the skin, muscle, fat, and liver, exposing the stomach more clearly. Next, a second cut was applied to a part of the liver and the spine to better reveal the heart (green dotted circle). In this way, different cuts were made with distinct objectives to achieve the desired visualization. 

\begin{figure*}[!t]
\centering
\subfloat[]{\includegraphics[width=1.725in]{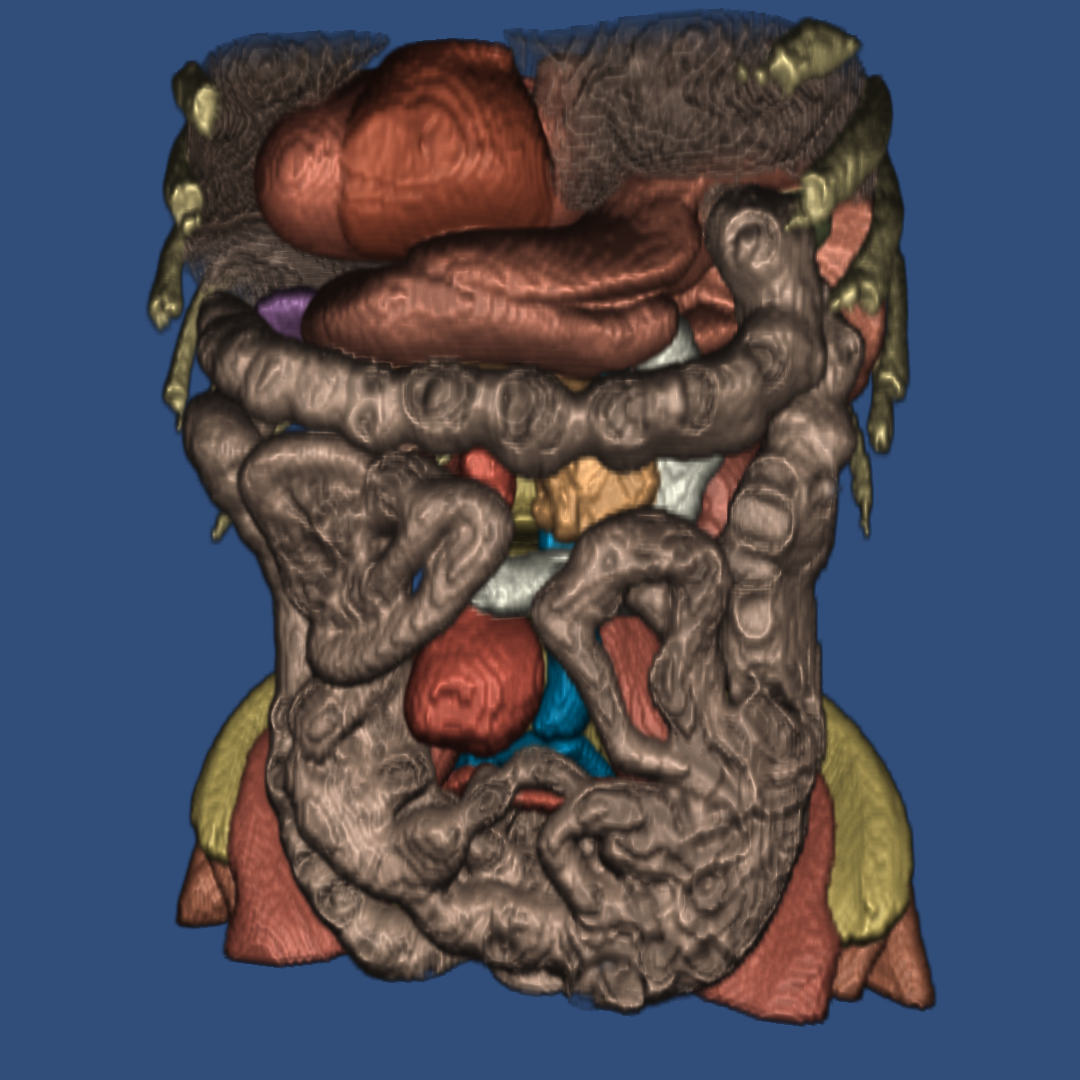}
\label{fig:steps_a}}
\hfil
\subfloat[]{\includegraphics[width=1.725in]{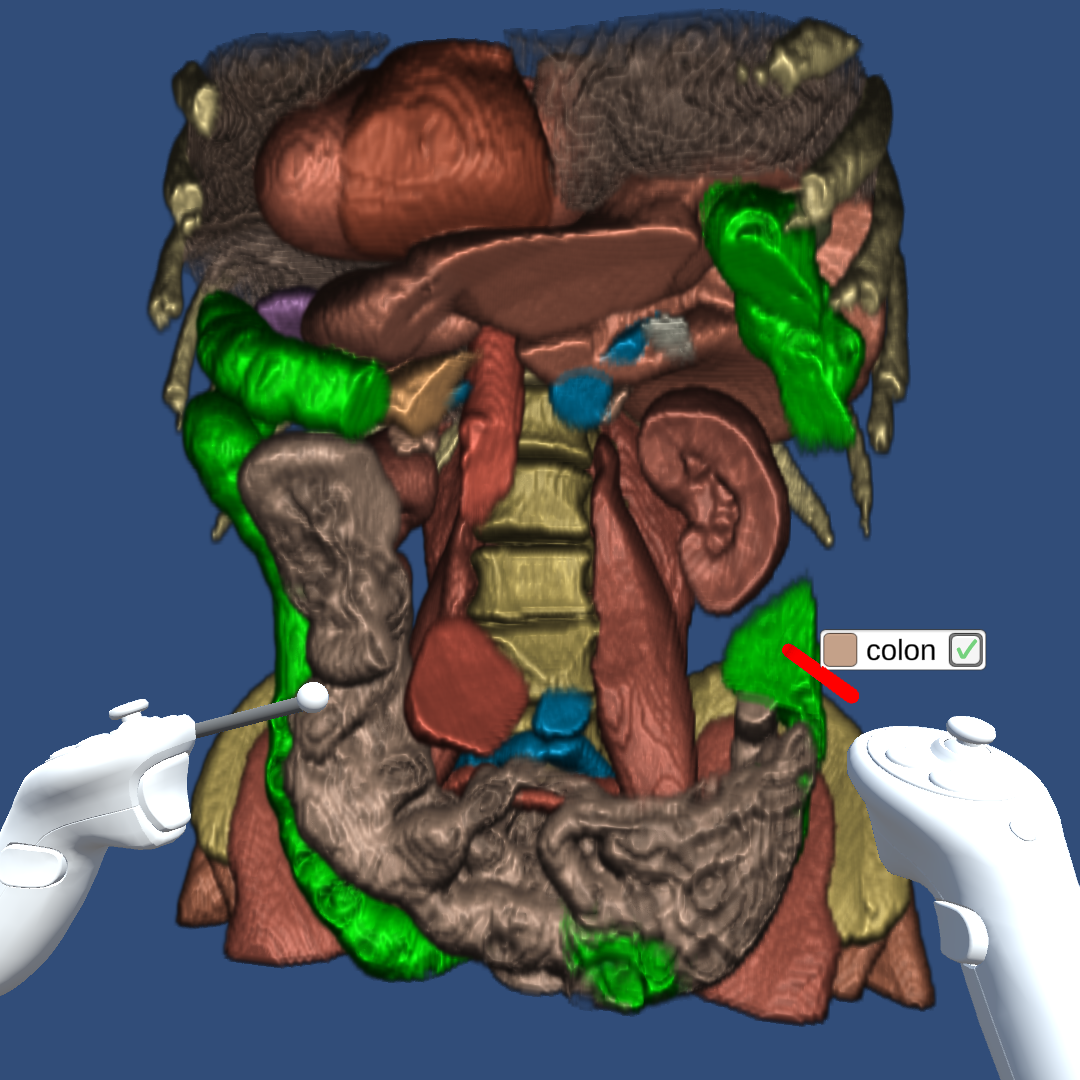}
\label{fig:steps_b}}
\hfil
\subfloat[]{\includegraphics[width=1.725in]{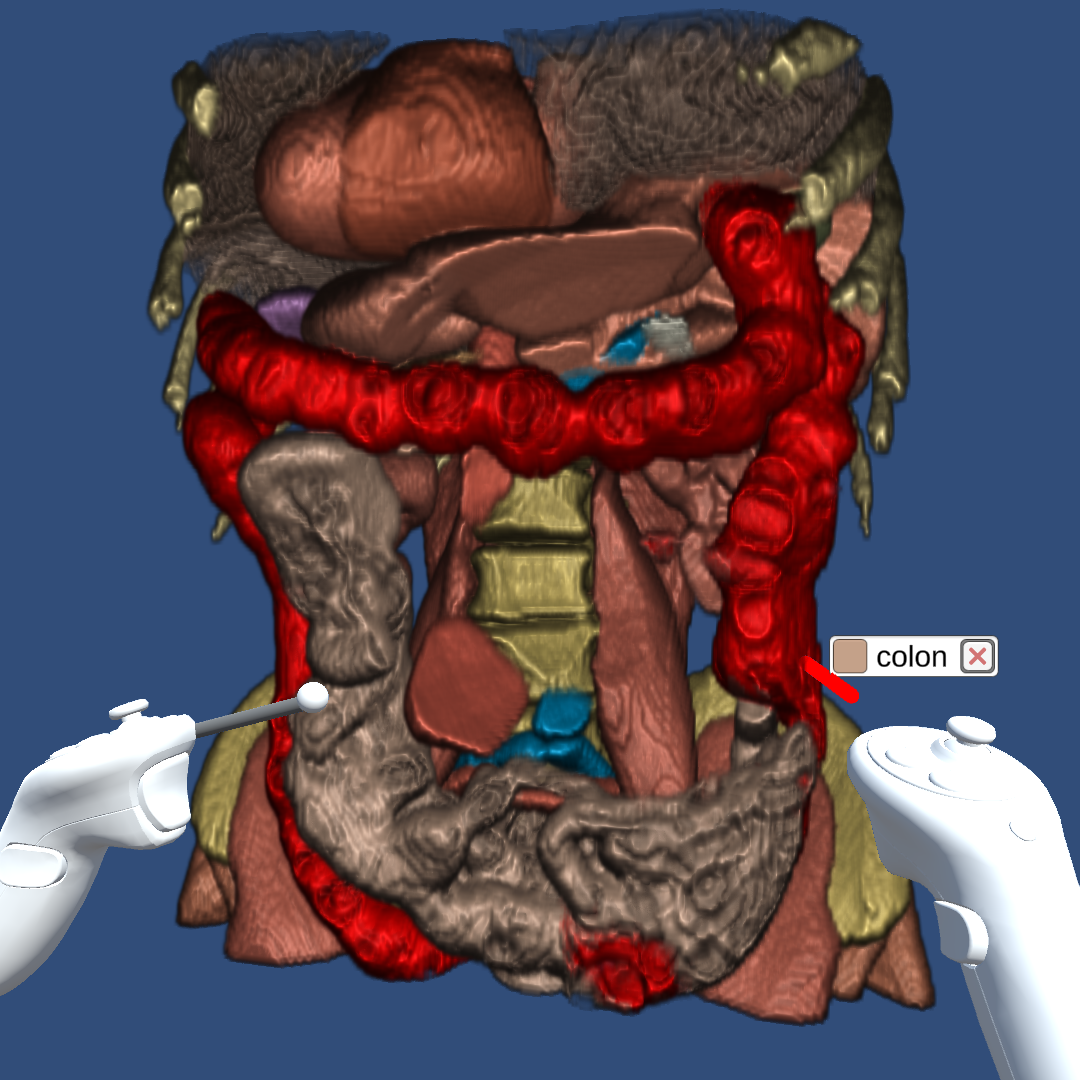}
\label{fig:steps_c}}
\hfil
\subfloat[]{\includegraphics[width=1.725in]{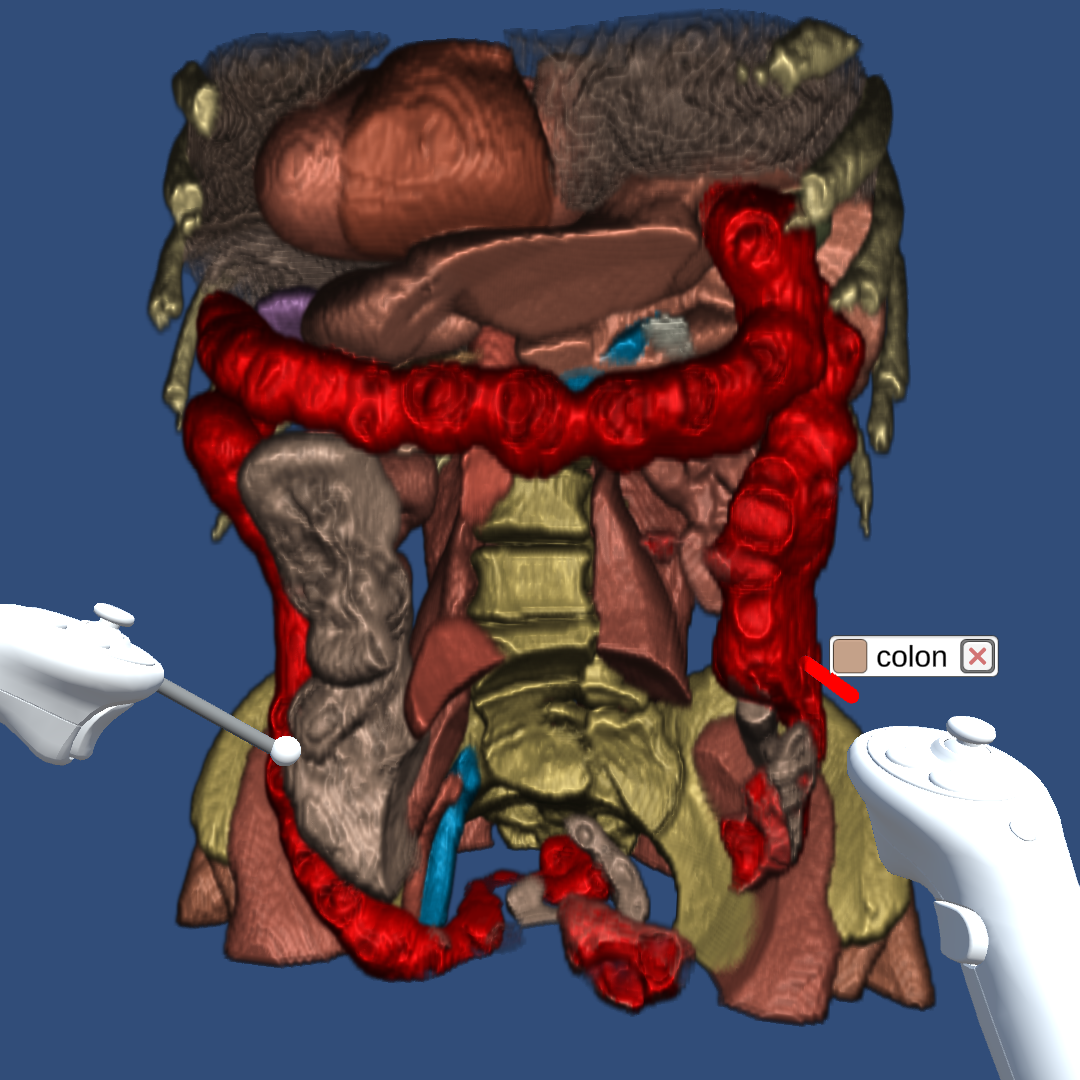}
\label{fig:steps_d}}
\caption{The basic functionality of AnatomyCarve. (a) A segmented medical volume where no clipping is applied is first shown. (b) A user inserts the clipping sphere in the volume, which hides all the voxels from all segments inside of it, and the user can simultaneously select segments on it. Then, in (c), the user makes the “colon” segment unclippable, so all of its voxels become visible, even if they are inside the clipping sphere. In (d), the user deposits the current clipping sphere and starts to manipulate a new one.}
\label{fig:steps}
\end{figure*}

\subsection{Interaction Paradigm}

AnatomyCarve employs an asymmetrical interaction model where each hand performs a different task. One hand is used to insert an invisible sphere into the volume, making all voxels within it invisible. The other hand is responsible for selecting which segments should be clipped by the invisible sphere and which should remain visible. A video demonstrating this functionality is provided along with the article.

AnatomyCarve allows the user to interact with a segmented volume (see Fig. \ref{fig:steps_a}) by selecting anatomical segments (e.g., colon, as shown in Fig. \ref{fig:steps_b}) among those that are present on the segmented volume. They can simultaneously insert a clipping sphere inside the volume, which hides the voxels within it, similar to the algorithm presented by Joshi et al. \cite{joshi_novel_2008} (see Fig. \ref{fig:steps_b}). However, segments marked as ``unclippable" by the user remain visible, with all of their voxels rendered, as shown in Fig. \ref{fig:steps_c}. The user can then fix the clipping sphere in place, which continues to clip voxels that are not marked as unclippable. New clipping shapes can be added to carve additional regions, with potentially different segments to clip, as demonstrated in Fig. \ref{fig:steps_d}.

To decide which segments are clipped, a classic laser pointer interaction (ray casting) is used. The user manipulates the pointer to target segments, which will change color to green or red to indicate whether the segment will be clipped. The user can toggle this state for any selected segment by pressing a button. The laser beam stops at the first visible segment and selects it. After selection, the segment's color changes to either green or red (see Fig. \ref{fig:steps_b} and Fig. \ref{fig:steps_c}), depending on whether the segment is clippable, as indicated by a mask variable. Initially, all segments are clippable (see Fig. \ref{fig:joshi-alike}). The user can toggle the mask value by pressing the trigger on their controller, changing the segment's color from red to green or vice versa. A secondary button on the controller resets the mask, enabling or disabling clipping for all segments within the current clipping sphere.

The controller for the other hand has an invisible sphere attached, which the user can insert into the volume to hide clippable voxels. The user can modify the size of the sphere using the joystick and ``fix" the sphere in the volume by pressing the secondary button on the controller. When fixed, the sphere's mask is saved, and it will no longer change. A new smaller clipping sphere is instantiated with a copy of the mask from the fixed sphere, which the user can then modify. This is illustrated in Fig. \ref{fig:sketch}, where the first clipping mesh (blue) is fixed in place, and a second clipping mesh (green) is being manipulated. This type of interaction allows the user to "stack" clipping spheres step by step in the volumetric image, where each clips a different set of segments. 

Users can also remove the last placed clipping sphere by pressing the primary button. Additionally, they can move and rotate the volume by pressing and holding the grip button on either controller. Pressing both grip buttons simultaneously allows for intuitive translation, rotation, and scaling of the volume using both hands, enabling users to position the volume comfortably in front of them for efficient clipping.

\subsection{Implementation}
AnatomyCarve was implemented using the Unity Engine (Unity Technologies, United States, San Francisco), and uses a customized volume rendering algorithm. To clip parts of the volume, a calculation is performed for each voxel on the GPU using a compute shader to determine if that voxel should be hidden. 

\begin{figure*}[htb]
\centering{\includegraphics[width=7in]{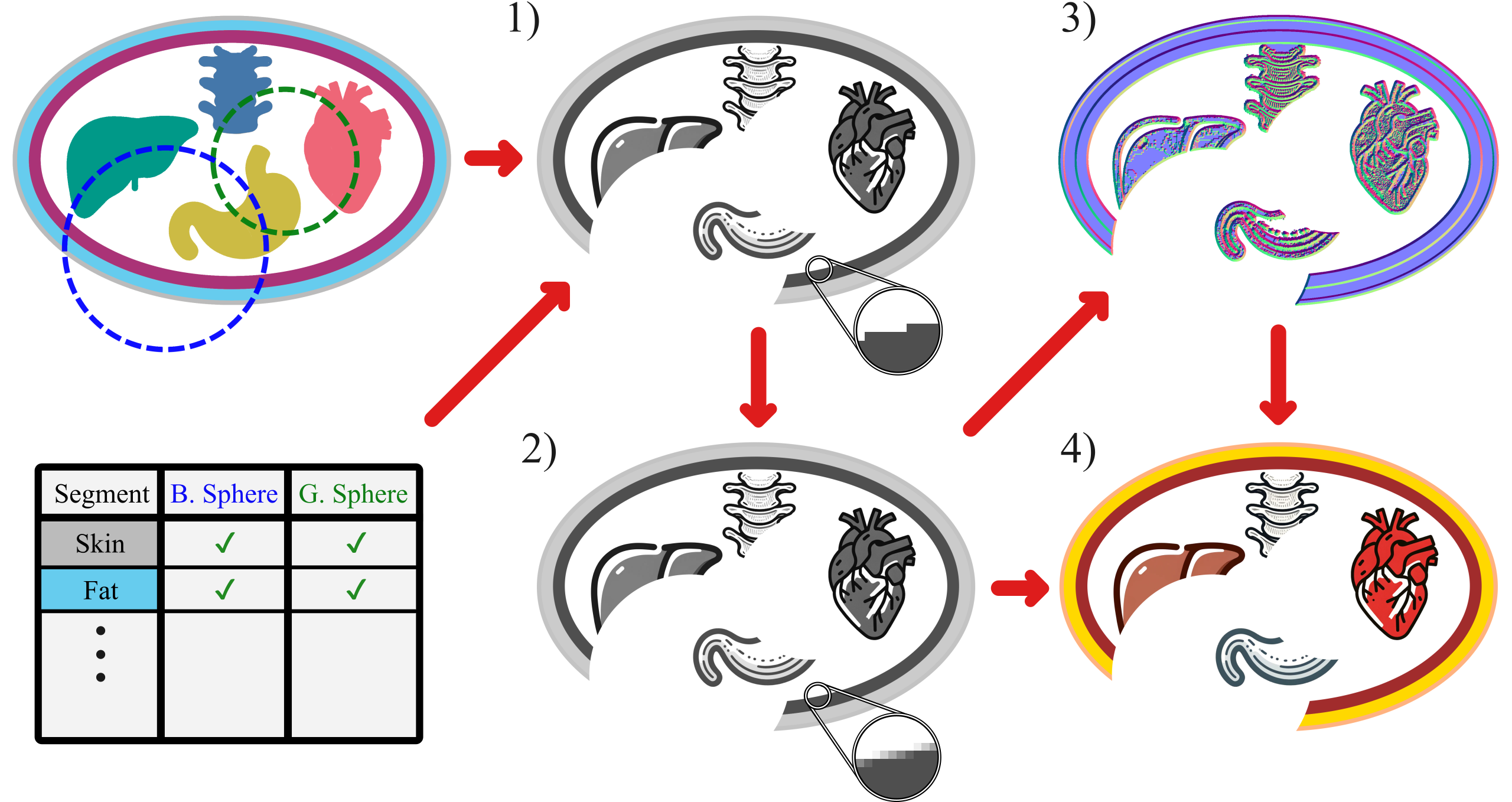}}
\caption{The volume rendering pipeline of AnatomyCarve, illustrated in 4 steps, each corresponding to a compute shader. In step (1), shadow test volume clipping is used to clip voxels from the specific segments using the clipping mask. In step (2), anti-aliasing is performed on the created opacity volume. In step (3), surface normals are calculated. Then, in step (4), the final rendering is performed.}
\label{fig:rendering-pipeline}
\end{figure*}

The algorithm requires the following inputs: an intensity volume that holds the original values from the CT or MRI scan, a 3D label map that associates each voxel in the intensity volume with a segment label, an opacity transfer function that maps intensity values to opacity, and a color transfer function that assigns a color to each segment label. For the color transfer function, we use the predefined color mapping from 3D Slicer \cite{fedorov_3d_2012}. The volume rendering process consists of four steps, each executed in a different compute shader (see Fig. \ref{fig:rendering-pipeline}):
\begin{enumerate} \item Shadow test volume clipping \cite{weiskopf_interactive_2003} is used to clip voxels from specific segments. For each voxel, we check if it is inside at least one clipping sphere, based on the clipping mask. If a voxel is inside at least one clipping sphere, it is marked as invisible; otherwise, it remains visible. This process can be formalized with the following equation:
\begin{equation}
  \label{eq:is_clipped}
  Clipped(v)=\bigvee_{i=1}^n Inside(s_i,v) \wedge Mask(s_i,label(v)),
\end{equation}

where $v$ represents the voxel with $(x,y,z)$ coordinates, $n$ is the total number of spheres in the scene, $s_i$ is the sphere at index $i$, $label(v)$ is the label of voxel $v$ from the label map, $Inside(s_i,v)$ checks if voxel $v$ is inside the sphere $s_i$, and $Mask(s_i, label(v))$ retrieves the clipping mask value for sphere $s_i$ and segment corresponding to $label(v)$. If the voxel is not clipped, we apply the opacity transfer function to it and store the result in the opacity volume. If the voxel is clipped, an opacity value of 0 is stored instead.

\item A compute shader takes the opacity volume as input and performs three-dimensional anti-aliasing (FXAA) \cite{lottes_timothy_fxaa_2009} on it, producing an anti-aliased opacity volume.

\item A Sobel operator is applied to the anti-aliased opacity volume to compute the surface normals, which are stored in a separate volume.

\item The final rendering is performed in the visible frame buffer using direct volume rendering. The anti-aliased opacity volume is used as the per-voxel opacity, and the normals volume is used for Blinn-Phong shading to enhance the visual quality.

\end{enumerate}

Optionally, we incorporated Local Ambient Occlusion (LAO) \cite{hernell_local_2010} during  clipping to enhance the final output display. However, LAO was not used during the evaluations due to slow performance during the interaction.

\section{User Studies}
We conducted two user studies to evaluate AnatomyCarve:
\begin{enumerate}
    \item A usability study with non-experts with the goal of evaluating how difficult it is for new users to create complex visualizations using the tool
    \item A study with neurosurgeons and neurosurgical residents to assess the usefulness of the system for a practical application, namely surgical planning.
\end{enumerate}
The following sections describe the details of those studies.

\subsection{Novice Usability Evaluation}

The usability study aimed to determine the difficulty for new users in obtaining a desired visualization using AnatomyCarve. We manually created a series of visualizations based on similar anatomical drawings and asked users to replicate the illustrations using AnatomyCarve. To assess the quality of their replicated visualization, we conducted a second study phase in which we asked participants to rate the quality of the replication compared to the model.  

For the design of this user study, we were inspired by the validation of the adaptive cross-sections algorithm by Díaz et al. \cite{diaz_adaptive_2012}, where participants had to use the tool to copy the images that were presented to them. Our rationale for conducting a study of this kind was that no other interactive occlusion management tool which works with segmented data allows the creation of such customizable illustrations without being constrained by the focus and context visualization paradigm. Thus, no fair contender technique could be found that wouldn't undermine the capabilities of AnatomyCarve.

The first part of the study was conducted in the laboratory using a VR headset running AnatomyCarve and the second part was conducted online after all participants completed the first part. 

\subsubsection{Data Preparation}
During the study, we displayed medical volumes from the CT-ORG dataset \cite{rister_ct-org_2020}, which were automatically segmented using TotalSegmentator \cite{wasserthal_totalsegmentator_2023,isensee_nnu-net_2021} in 3D Slicer \cite{fedorov_3d_2012}.  

Using this dataset, we manually created 11 distinct 2D views using AnatomyCarve. These views depicted anatomical images composed of multiple structures, with specific regions strategically removed to reveal hidden anatomical structures. On average to create the visualizations we inserted \(3.9 \pm 1.2\) (mean $\pm$ SD) spheres in the 8 non-trial volumes, and toggled the clippability of the segments \(25.8 \pm 28.8\) (mean $\pm$ SD) times. Screenshots were taken of all the created views, which were later presented as ground truth visualizations to be recreated by participants. 

\subsubsection{Pre-tests Questionnaire and Tutorial}
The user study began with informed consent, followed by a pre-test questionnaire querying participants about their experience with VR and anatomical medical images. Next, participants did a tutorial, which included a short presentation explaining what AnatomyCarve is and how it can be used. Afterward, participants put on the VR headset, and we guided them through the system's various interactions, step by step, to help them learn how to use it effectively.

\subsubsection{Task 1: Illustrative Recreation}
\begin{figure}[!t]
\centering
\subfloat[]{\includegraphics[width=1.725in]{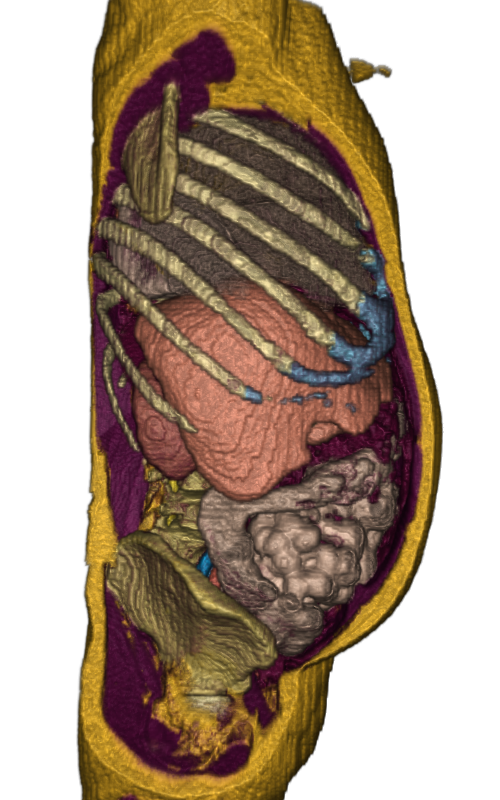}%
\label{fig:firstlast_a}}
\hfil
\subfloat[]{\includegraphics[width=1.725in]{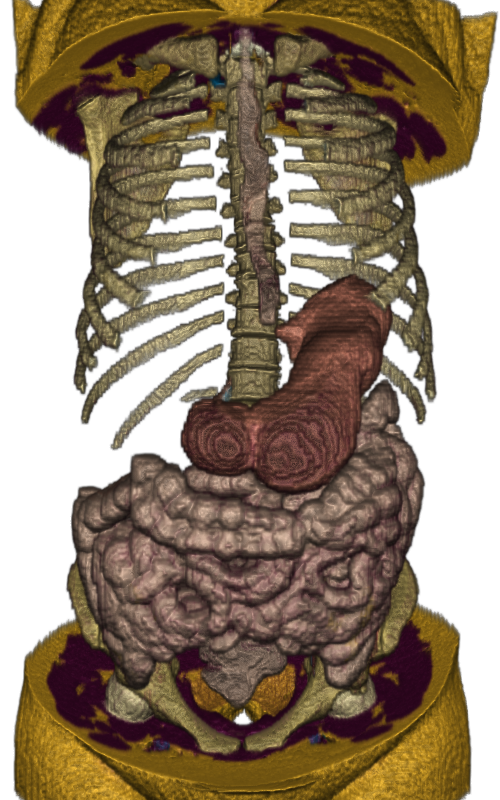}%
\label{fig:firstlast_b}}
\caption{The first (a) and the last (b) volumes that participants had to copy during the usability user study.}
\label{fig:firstlast}
\end{figure}

After completing the tutorial and becoming familiar with the controls, participants began Task 1. In each trial, they were presented with an unaltered medical image (i.e., with no voxels removed) and asked to use AnatomyCarve to remove the correct voxels from the appropriate segments, in order to make the 3D volume resemble the 2D ground truth view that was displayed. Participants had to ensure that the recreated volume matched the ground truth view when viewed from the same perspective. Once satisfied with their result, they pressed a button in the scene to proceed to the next trial. The user study included 3 guided test trials followed by 8 recorded trials that got progressively more difficult (requiring more inserted meshes and/or more segments to invert). During the test trials, participants received assistance to ensure they understood the system. For the recorded trials, they used AnatomyCarve independently, though they were free to ask questions if needed. Fig. \ref{fig:firstlast} shows the first and last (8th) volume that participants were asked to replicate during the recorded portion of the study.

\subsubsection{Task 2: Illustration Ranking}

In Task 2, our goal was to subjectively evaluate the quality of the visualizations created with AnatomyCarve in VR. After all participants completed the first part of the user study, we sent out an online questionnaire. In this questionnaire, each participant was shown the 8 original ground truth views they had been asked to replicate in Task 1, along with 7 anonymized participant recreations for each view. Their task was to rank (i.e., reorder) the 7 recreations from most to least similar to the original ground truth view. Each participant evaluated only a subset of the total views. Specifically, for each of the 8 trials, every participant was assigned 7 out of the 17 total recreations, chosen in a pseudo-random but balanced way so that each recreation was evaluated exactly 7 times across all participants. All views were presented anonymously—both to protect participant privacy and to avoid bias in the rankings.

\subsubsection{Evaluation Measures}

In order to evaluate the usability of the system, we used the following metrics and questionnaires for Task 1:

\begin{itemize}
  \item Time per trial.
  \item Ratio of inserted clipping spheres by the users compared to the number that we have inserted when creating the ground truth views.
  \item Ratio of toggled segments by the users compared to the number that we have toggled when creating the ground truth views.  
  \item System Usability Scale (SUS) \cite{brooke_sus_1996}.
  \item NASA Task Load Index (TLX) \cite{hart_development_1988}.
  
\end{itemize}

\noindent For Task 2, we measured: 
\begin{itemize}
  \item The root mean squared error (RMSE) of the depth for each created view, which measures how different is the depth map of the participant-created view image and the ground truth view image. 
  \item The mean absolute error (MAE) of the first encountered segment for each created view, which measures, for each corresponding pixel of the user-created view, if the first visible segment that was encountered during ray casting is the same or not as in the ground truth view. It is a new metric that we introduce as a contender for the RMSE of depth metric for comparing segmented medical images.
  \item The subjective global rank of each view, rated by the participants, calculated using the Plackett-Luce model \cite{plackett_analysis_1975}.
\end{itemize}

\begin{figure*}[!t]
\centering
\subfloat[]{\includegraphics[width=1.725in]{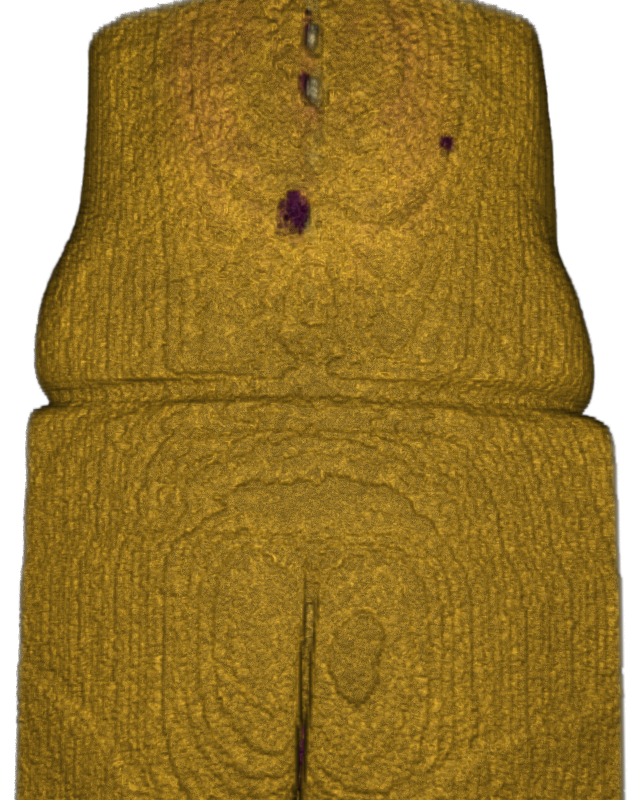}
\label{fig:volume85_a}}
\hfil
\subfloat[]{\includegraphics[width=1.725in]{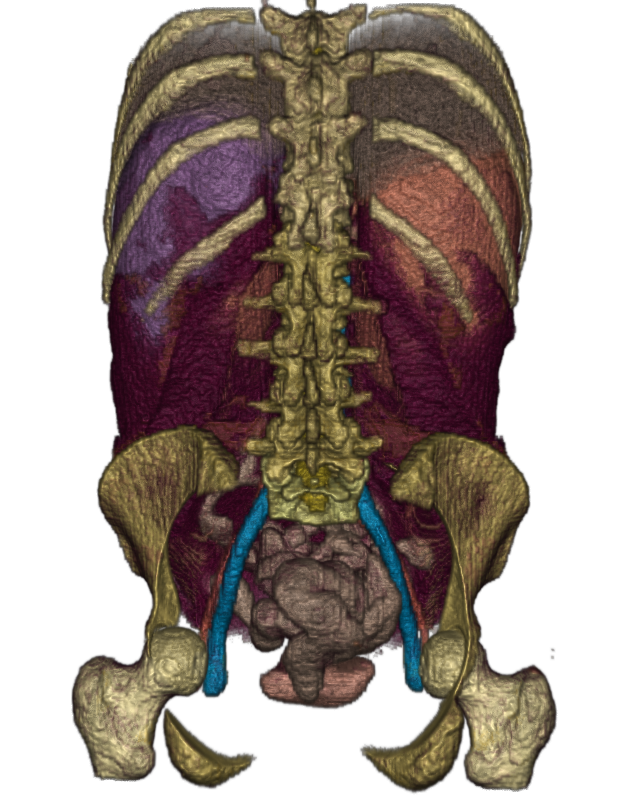}
\label{fig:volume85_b}}
\hfil
\subfloat[]{\includegraphics[width=1.725in]{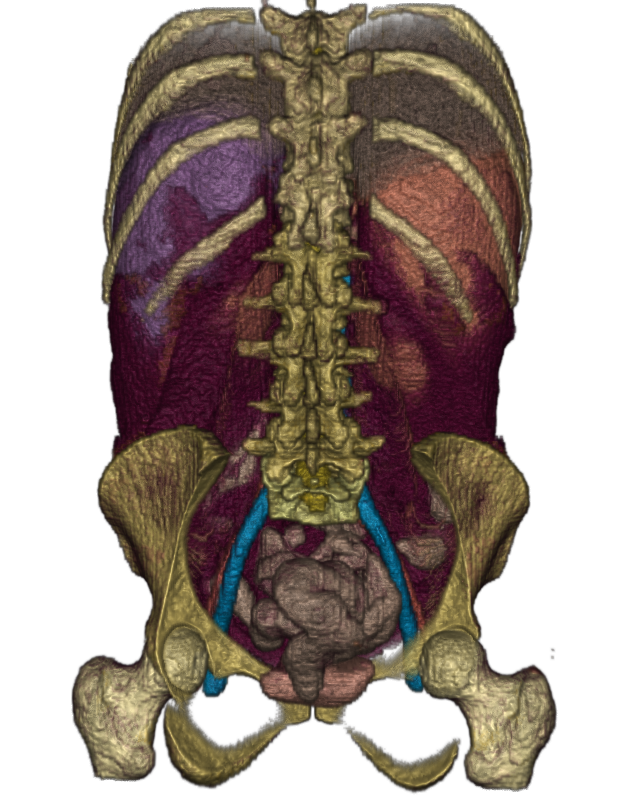}
\label{fig:volume85_c}}
\hfil
\subfloat[]{\includegraphics[width=1.725in]{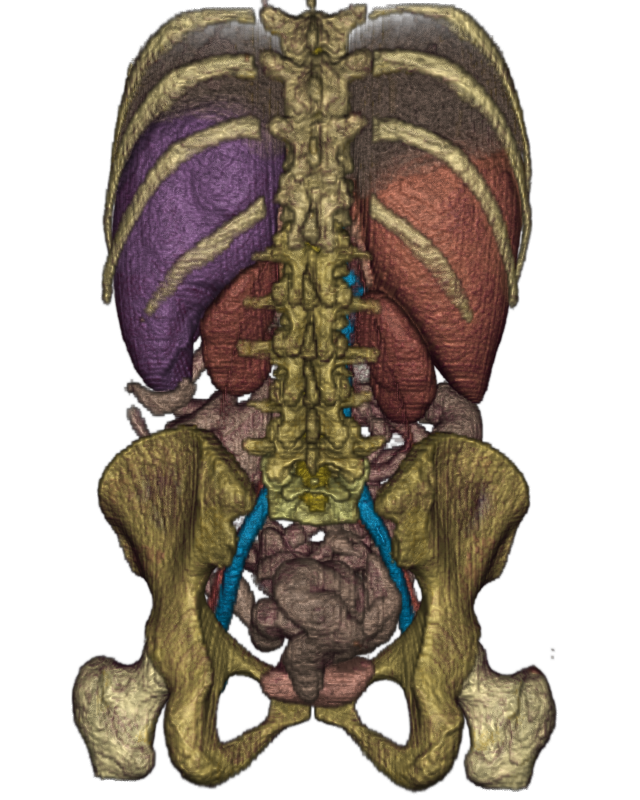}
\label{fig:volume85_d}}
\caption{The volume that was presented at the 5th trial to the participants. (a) represents the original volume that was presented to the participant, (b) represents the ground truth image that was created by the researchers and that all the participants tried to copy, (c) represents the image create by the participant at sextile rank S1 (3rd best view out of 17), and (d) represents the image created by the participant at sextile rank S5 (15th best view out of 17).}
\label{fig:volume85}
\end{figure*}

\begin{figure*}[!t]
\centering
\subfloat[]{\includegraphics[width=1.725in]{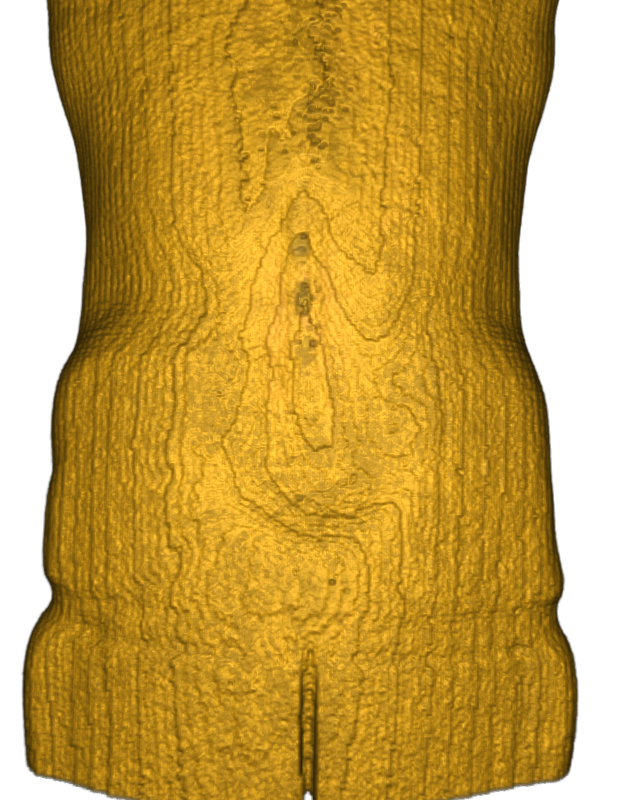}
\label{fig:volume105_a}}
\hfil
\subfloat[]{\includegraphics[width=1.725in]{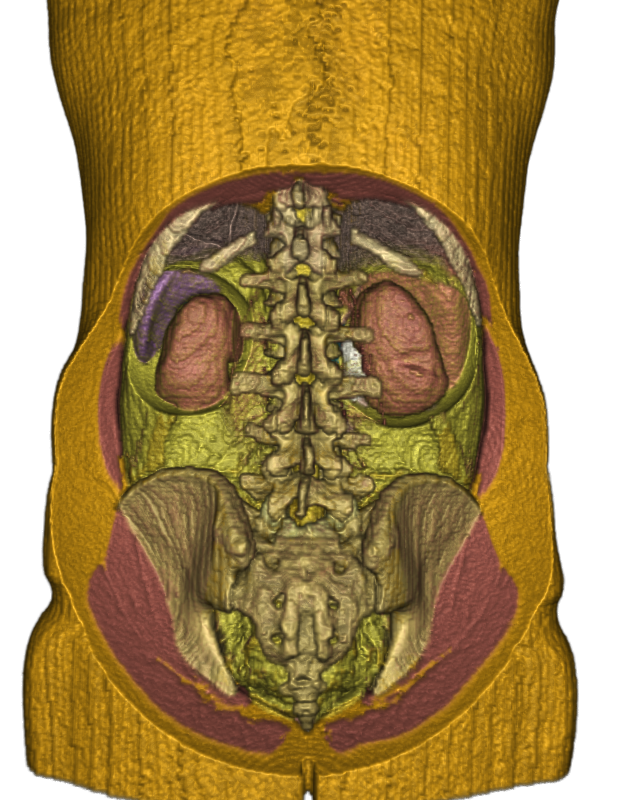}
\label{fig:volume105_b}}
\hfil
\subfloat[]{\includegraphics[width=1.725in]{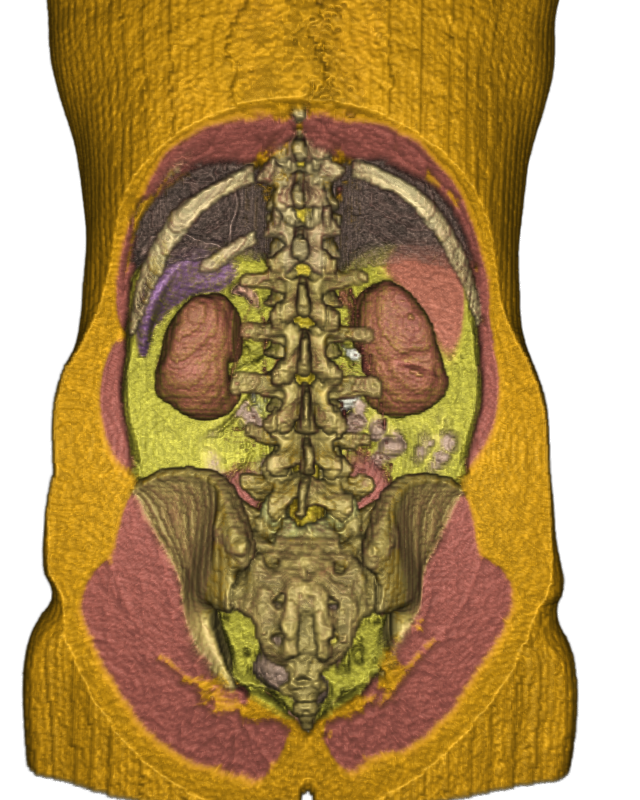}
\label{fig:volume105_c}}
\hfil
\subfloat[]{\includegraphics[width=1.725in]{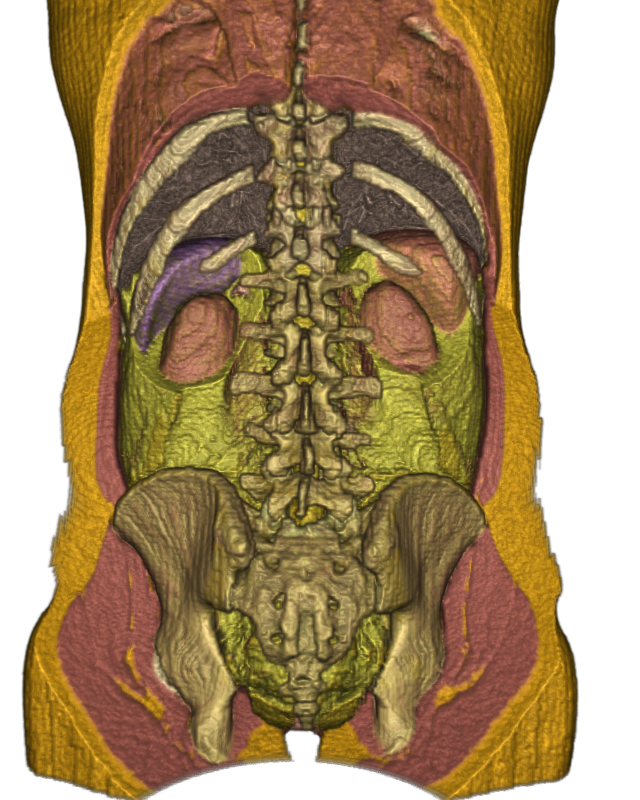}
\label{fig:volume105_d}}
\caption{The volume that was presented at the 7th trial to the participants. (a) represents the original volume that was presented to the participant, (b) represents the ground truth image that was created by the researchers and that all the participants tried to copy, (c) represents the image created by the participant at sextile rank S1 (3rd best view out of 17), and (d) represents the image created by the participant at sextile rank S5 (15th best view out of 17).}
\label{fig:volume105}
\end{figure*}

\begin{figure*}[!t]
\centering
\subfloat[]{\includegraphics[width=1.725in]{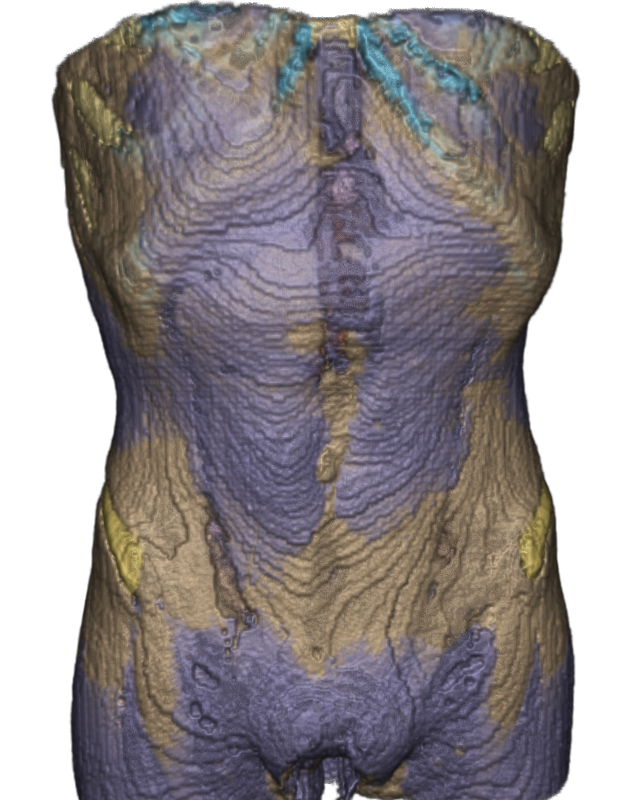}
\label{fig:volume130_a}}
\hfil
\subfloat[]{\includegraphics[width=1.725in]{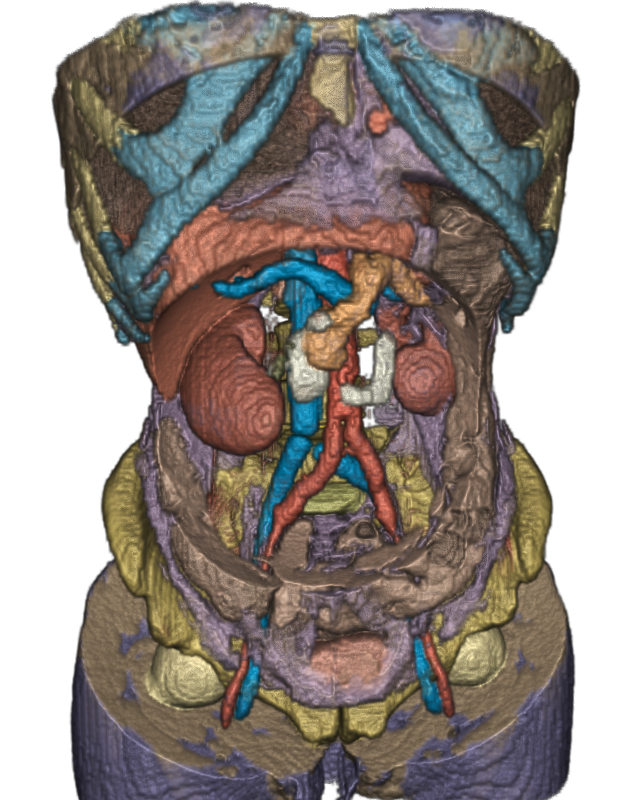}
\label{fig:volume130_b}}
\hfil
\subfloat[]{\includegraphics[width=1.725in]{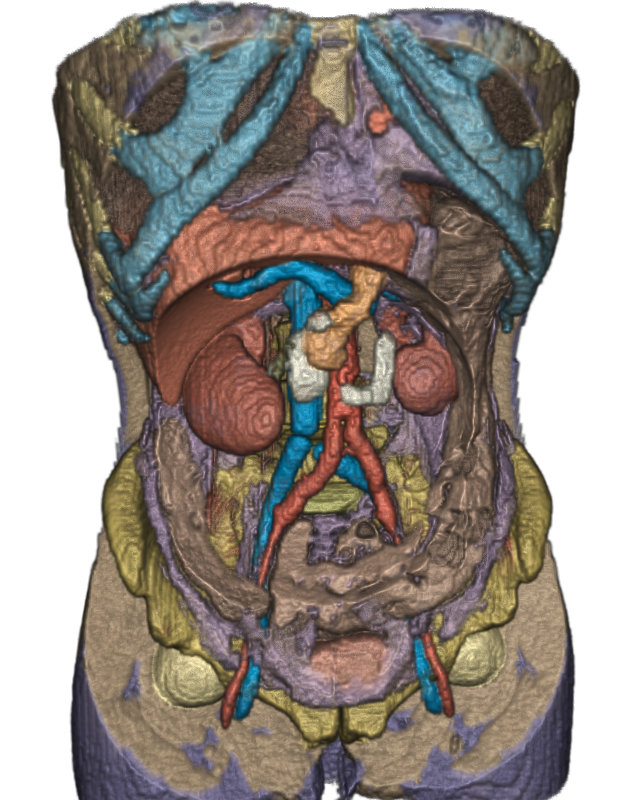}
\label{fig:volume130_c}}
\hfil
\subfloat[]{\includegraphics[width=1.725in]{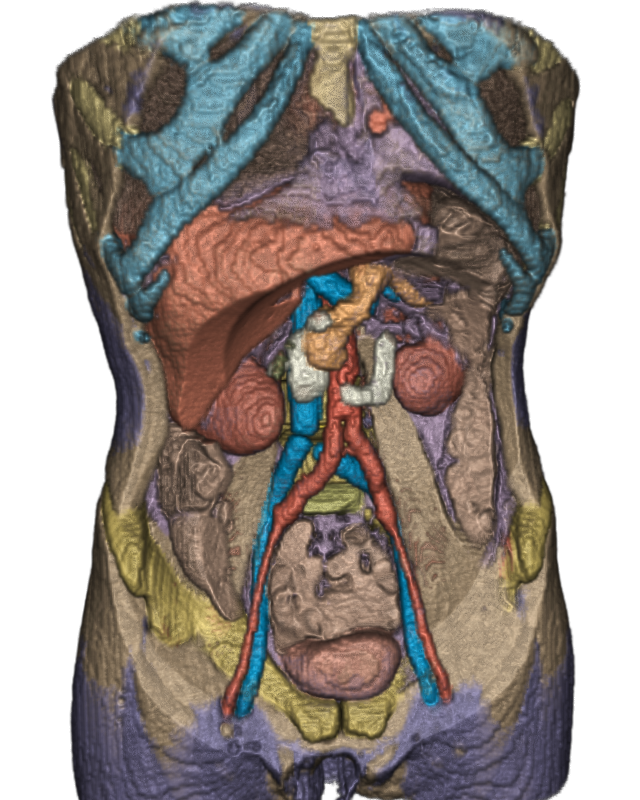}
\label{fig:volume130_d}}
\caption{The volume that was presented at the 3rd trial to the participants. (a) represents the original volume that was presented to the participant, (b) represents the ground truth image that was created by the researchers and that all the participants tried to copy, (c) represents the image created by the participant at sextile rank S1 (3rd best view out of 17), and (d) represents the image created by the participant at sextile rank S5 (15th best view out of 17).}
\label{fig:volume130}
\end{figure*}

\subsection{Expert Evaluation}
The expert user study focused on surgical planning for the removal of tumors from the spine. A total of six volumes were selected from former patients at Anonymous Hospital: three with tumors in the thoracic spine, two in the cervical spine, and one in the intramedullary spine.

\subsubsection{Data Preparation}

Anatomical segmentation of bones and soft tissues was performed using TotalSegmentator \cite{wasserthal_totalsegmentator_2023,isensee_nnu-net_2021}, followed by manual refinement and validation by a medical doctor. Vascular structures were segmented semi-automatically using the “GrowCut” tool in the open-source application 3D Slicer \cite{kikinis_3d_2011}, also by the same expert.

\subsubsection{Pre-test Questionnaire and Tutorial}
Prior to the study, the experts gave their consent and completed a pre-test questionnaire about their familiarity with VR and clinical experience. Next, they were given a tutorial on how to use the system. 

\subsubsection{Task}

The primary objective of the expert study was to simulate traditional surgical planning. Participants used AnatomyCarve to plan trajectories for tumor resection. This involved (1) identifying the tumor location within the segmented spine volume; (2) determining a safe and effective trajectory to access the tumor and (3) identifying anatomical structures that would need to be traversed or avoided during the procedure. 

\subsubsection{Evaluation Measures}

To assess the usability and perceived effectiveness of the tool, participants were asked to complete the SUS questionnaire \cite{brooke_sus_1996} at the end of the session. Additionally, they were invited to provide qualitative feedback on the strengths and weaknesses of the system, limitations encountered during use, and suggestions for future improvements.

\subsection{Apparatus} 
The studies were conducted on an Oculus Quest 2 headset (Meta Platforms, Inc., United States, Menlo Park). To ensure smooth rendering of the volumetric data, the Oculus Quest 2 was set to a refresh rate of 72 Hz. For the novice study we used a Windows 11 machine with an Intel Core i9-11900H processor, 32 GB of RAM, and an external AMD Radeon RX 6800 XT graphics card with 16 GB of VRAM. For the expert user study, we used a Windows 10 machine with an AMD Ryzen 7 3700X processor, 32 GB of RAM, and an NVidia RTX 2080 Ti graphics card with 11 GB of VRAM. Ethical approval for the user studies was obtained from the École de Technologie Supérieure ethics committee.

\section{Results}

\subsection{Novice Usability Evaluation}

\subsubsection{Participants}

A total of 17 non-expert participants took part in the study (8 identifying as male, 8 as female, and 1 as non-binary), with ages ranging from 23 to 36 years (\(M = 27.4, SD = 3.4\)). Participants were first asked about their familiarity with relevant domains. Overall, 82\% reported having at least moderate VR experience, with 41\% describing themselves as extremely experienced. Regarding anatomical medical images, 41\% of participants indicated moderate familiarity, while 53\% reported being only slightly familiar or not familiar at all.

\subsubsection{Task 1: Performance Metrics}

During the first part of the study, participants were asked to recreate anatomical views using \textit{AnatomyCarve}. On average, the participants were quick to recreate the anatomical visualizations, taking on average \(176 \pm 58\) seconds (mean $\pm$ SD) to recreate a single view. The average ratio of the clipping meshes inserted by the participants to the number originally inserted on the model was \(1.88 \pm 0.55\). Thus, users inserted nearly twice as many clipping shapes as the reference solution, suggesting they explored more or used smaller, more incremental edits. The toggling of segments closely matched the original, showing that users generally selected the appropriate structures. Specifically, the ratio of segment toggles performed by participants compared to the original was \(0.98 \pm 0.62\). 

\subsubsection{Task 1: SUS \& NASA TLX}

In terms of SUS, novices rated the system highly (77.6), which is above the average score of 68 \cite{bangor_empirical_2008}, indicating they found it relatively easy to use with minimal perceived complexity. They were confident in their ability to learn and use the system quickly, and did not feel the need for significant technical support (see Table \ref{tab:sus}).

\begin{figure}[htb]
\centering{\includegraphics[width=3.5in]{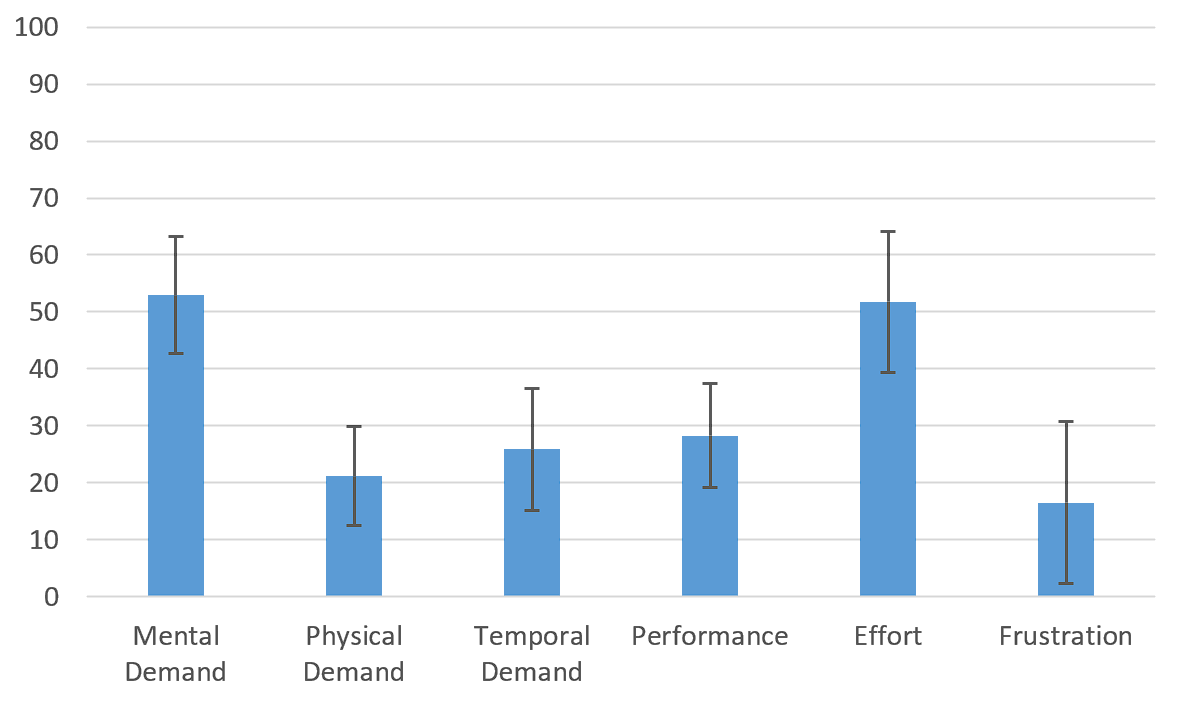}}
\caption{NASA TLX score for each of the 6 metrics, averaged for all novice participants. The error bars represent the standard deviation.}
\label{fig:nasa-tlx}
\end{figure}

The NASA-TLX resulted in an average workload score of 32.7. Results indicate that users found the task mentally demanding and requiring effort, with both dimensions scoring just above 50. However, frustration, physical demand, and temporal demand were rated relatively low, suggesting that users found the system usable and not overly taxing. Overall, participants were able to complete the task without significant stress or discomfort as can be seen in Fig. \ref{fig:nasa-tlx}.

\begin{table}[!t]
\caption{SUS Results of the two user studies.\label{tab:sus}}
\centering
\begin{tabular}{|p{4.4cm}|p{1.3cm}|p{1.3cm}|}
\hline
\textbf{Question} & \textbf{Novices} & \textbf{Experts} \\
\hline
I think that I would like to use this system frequently. & $4.06\pm0.73$ & $4.13\pm0.78$ \\
\hline
I found the system unnecessarily complex. & $1.76\pm0.64$ & $1.63\pm0.48$ \\
\hline
I thought the system was easy to use. & $4.24\pm0.73$ & $4.38\pm0.86$ \\
\hline
I think that I would need the support of a technical person to be able to use this system. & $2.53\pm1.14$ & $3.13\pm0.78$ \\
\hline
I found the various functions in this system were well integrated. & $4.35\pm0.76$ & $4.25\pm0.66$ \\
\hline
I thought there was too much inconsistency in this system. & $1.12\pm0.32$ & $1.75\pm0.66$ \\
\hline
I would imagine that most people would learn to use this system very quickly. & $3.82\pm0.92$ & $4.13\pm0.78$ \\
\hline
I found the system very cumbersome to use. & $1.71\pm0.82$ & $2.00\pm0.71$ \\
\hline
I felt very confident using the system. & $3.65\pm0.97$ & $4.13\pm0.93$ \\
\hline
I needed to learn a lot of things before I could get going with this system. & $1.94\pm0.87$ & $2.38\pm1.11$ \\
\hline
\textbf{Total Score} & \textbf{77.6} & \textbf{75.3}\\
\hline
\end{tabular}
\end{table}

\subsubsection{Task 2: Illustration Ranking}

\begin{figure}[!ht]
\centering
\includegraphics[width=3.5in]{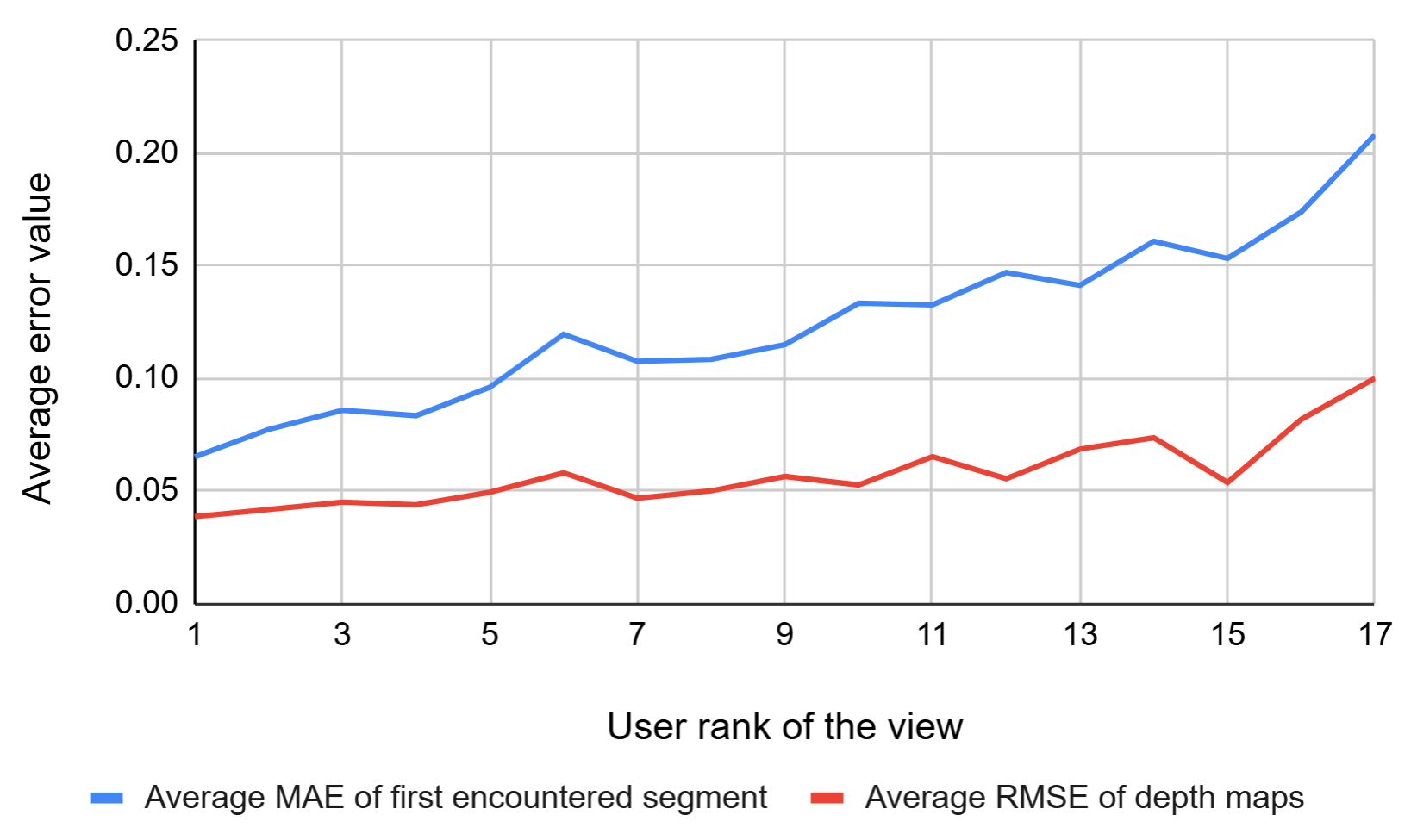}
\caption{Average MAE of the first encountered segment for each pixel, plotted by rank and averaged across all trials.}
\label{fig:mae-per-rank}
\end{figure}

In the second part of the study, participants were asked to rank recreated views based on their similarity to the original ground truth views. These rankings were aggregated using the Plackett-Luce model~\cite{plackett_analysis_1975} to generate a global ranking of all participant-created views for each trial. We also calculated two objective similarity metrics between the user-created views and the ground truth: root mean square error (RMSE) of depth and mean absolute error (MAE) of the first encountered segment. Fig. \ref{fig:mae-per-rank} shows the relationship between participant rankings and these metrics, averaged over all 8 trials. The trend line for average MAE of the first encountered segment achieved an \(R^2\) of 0.93 with a slope of 0.059, while the RMSE trend line had an \(R^2\) of 0.70 and a slope of 0.034.

We also computed average MAE and RMSE values across all participants for each trial. The volume with the lowest average MAE of the first encountered segment (\(0.066 \pm 0.040\)) was from the 5th trial, shown in Fig. \ref{fig:volume85}. In contrast, the volume from the 7th trial (Fig. \ref{fig:volume105}) had the highest error (\(0.167 \pm 0.058\)). Notably, the volume from the 3rd trial (Fig. \ref{fig:volume130}) had the steepest slope and highest \(R^2\) values when mapping user rankings to both MAE and RMSE metrics (\(R^2 = 0.84\) for MAE, \(R^2 = 0.625\) for RMSE). This indicates a large disparity in performance between participants, suggesting that this task was more polarizing in terms of difficulty.

\subsection{Expert Evaluation}

\subsubsection{Participants}
The usefulness of AnatomyCarve for surgical planning was evaluated by 8 practicing neurosurgeons or neurosurgical residents (6 identifying as male, 2 as female) from Anonymous. The participants were between 26-46 years old (\(M = 34.9, SD = 6.9\)). Participants had on average \(7.4 \pm 5.6\) (mean $\pm$ SD) years of clinical experience. We also determined that 3 out of 8 participants reported using VR at least once per month, while 5 out of 8 said they had never used VR.

\subsubsection{SUS}
In terms of SUS scores, experts rated the system slightly lower (75.3) compared to novices (77.6). However, experts generally reported higher confidence in using the system, finding it easier to use and easier to learn, although they did perceive the system to be slightly more inconsistent and requiring more technical support than novices did (see Fig. \ref{tab:sus}). 

\subsubsection{Expert Feedback}
In terms of user feedback, a majority of participants (7 out of 8) appreciated the three-dimensional view of anatomical structures created by the system, which they felt was a significant improvement over traditional surgical planning methods. Traditional methods typically provide only 2D views, such as axis-aligned slicing (axial, coronal, sagittal), or simple 3D rendered volumetric representations. Specifically, 4 out of 8 participants highlighted that the 3D visualization in AnatomyCarve was especially useful for preoperative evaluation and planning. Moreover, 4 out of 8 participants noted that the 3D visualization helped clarify the spatial arrangement and connections between anatomical features, which improved their understanding of anatomical relationships during surgical planning.

In terms of weaknesses, 5 out of 8 participants mentioned that the resolution and/or precision of the visualized volumes were insufficient, making it difficult to identify finer anatomical structures. Additionally, 3 out of 8 participants noted that the absence of 2D imaging options made it harder to mentally map traditional imaging modalities to the 3D views. Another 3 participants pointed out that the system had a learning curve, suggesting that it might take some time for users to become proficient.

Finally, in terms of suggestions for improvement, 5 out of 8 participants recommended enhancing the system's resolution to allow for better visualization of smaller anatomical details. Further, 4 out of 8 participants suggested adding the ability to view 2D images alongside the 3D visualization to improve usability. One surgeon also proposed adding an eraser-like tool to continuously clip elements from different segments, which would improve the system's flexibility during surgical planning.

\section{Discussion}

The evaluation of \textit{AnatomyCarve} has demonstrated positive results both from non-expert and expert participants, with findings that suggest the system is effective and intuitive for creating anatomical visualizations and effective for surgical planning, though there are some areas for improvement. 

In terms of usability, both non-expert and expert participants provided similar SUS scores, 77.6 and 75.3, respectively, indicating that the interface was generally intuitive and AnatomyCarve was easy to use. The NASA TLX score of 32.7 for non-expert participants suggests that, while the system is usable, there is a learning curve, which echoes the feedback from experts. We hypothesize that this learning curve was not solely due to the significant differences in visualization when compared to traditional modalities, but rather stemmed from the complexity of the system's interface, which requires remembering a high number of button actions and gestures.

An important insight from the second part of the usability study is the usefulness of MAE at the first encountered segment as a reliable objective metric for ranking visualization similarity. The higher \( R^2 \) and slope of the MAE trend line, compared to the RMSE, demonstrated that MAE provides a better reflection of the visualization’s fidelity. Specifically, the volume with the lowest MAE, shown in Fig. \ref{fig:volume85}, indicated a highly accurate recreation of the ground truth, where the key anatomical structures were clearly depicted. Conversely, the volume with the largest MAE, displayed in Fig. \ref{fig:volume105}, still showed critical anatomical features, such as the kidneys and pelvis, albeit with some missing details (e.g., skin and exposed muscles). Even in the worst-ranked volumes, the users were able to represent the majority of the correct segments, suggesting that AnatomyCarve is both simple to use and effective in producing accurate anatomical visualizations, even when non-experts with limited anatomical experience were involved.

The neurosurgeons and neurosurgical residents expressed a high level of satisfaction with AnatomyCarve’s 3D visualization capabilities. They particularly appreciated the ability to view spatial relationships between anatomical structures, a feature that is often difficult to achieve with traditional 2D imaging modalities. However, experts also identified some limitations, notably the resolution and precision of the visualized volumes, which made it difficult to clearly identify smaller structures such as nerves. This issue was not a limitation of the  AnatomyCarve system itself, but rather a result of the quality of the underlying dataset, which lacked the fine-grained segmentation needed to visualize such small anatomical structures. Moving forward, improving the resolution of the dataset and utilizing more precise segmentation techniques, possibly through automatic or semi-automatic methods, could help address this concern.

Another important area for improvement identified by experts was the system’s learning curve. A significant proportion of the expert participants suggested integrating 2D imaging alongside the 3D visualization to facilitate easier interpretation, especially for those more familiar with traditional 2D surgical planning tools. Incorporating 2D images into the VR scene could enable users to cross-reference traditional methods with the novel 3D views provided by AnatomyCarve. This integration could help reduce the learning curve and make the system more intuitive for surgeons and neurosurgical residents, ultimately improving its usability in a clinical setting.

Overall, AnatomyCarve demonstrates considerable potential as a tool for preoperative planning, providing quick and high-quality 3D anatomical visualizations. With further refinement of the system, including improved resolution, better segmentation, and the integration of 2D images, AnatomyCarve could become an even more valuable tool for surgical planning, especially in the neurosurgical field.

\section{Conclusion}

In this paper, we presented AnatomyCarve, a novel interactive occlusion management tool designed to enhance the visualization of segmented medical images in VR. The technique allows the viewer to select specific segments to clip, enabling the creation of complex visualizations akin to those found in anatomy textbooks. By combining the advantages of both clipping and deformation techniques, AnatomyCarve allows the viewer to perceive hidden 3D anatomical structures, while preserving the positions and spatial relations between them. 

The results of our qualitative evaluation demonstrate that AnatomyCarve can be used to create volumetric images such as those found in anatomical textbooks, while providing an intuitive and immersive experience. The expert evaluation suggests that AnatomyCarve has great potential for supporting surgeons in preoperative planning, as it was positively reviewed by practicing neurosurgeons and neurosurgical residents. Additionally, our usability study with non-experts confirmed that AnatomyCarve enables the creation of high-quality custom anatomical images in just a few minutes, making it an effective interface for interacting with medical data.

While AnatomyCarve was well-received, future work is needed to formally evaluate its effectiveness in comparison to traditional preoperative planning methods. Further exploration could focus on adding the ability to deform segments, simulating additional occlusion management techniques used in anatomical textbooks. One potential feature, suggested by one of the surgeons, is the inclusion of a continuous "eraser"-like tool. Unlike the current system, which only supports discrete clipping shapes, this tool would allow users to continuously insert clipping shapes while a button is pressed, simplifying the creation of custom anatomical visualizations. Additionally, improvements in rendering could be achieved by adapting more complex rendering methods, such as Contextual Ambient Occlusion \cite{titov_contextual_2024} to support interactive clipping and enable real-time enhancement of visualizations during the clipping process, rather than only for the final output.

In summary, AnatomyCarve represents an innovative approach to medical visualization, transforming traditional 2D anatomical illustrations into immersive 3D experiences. This 3D visualization not only enhances preoperative planning but also provides a powerful tool for medical education, enabling users to interact with anatomical structures in ways that static 2D images  or traditional clipping techniques cannot. By enabling the creation of highly customizable, interactive 3D views of medical data, AnatomyCarve allows users to explore complex anatomical relationships and gain a deeper understanding of the structures involved. With further improvements, such as incorporating continuous clipping tools and real-time rendering enhancements, AnatomyCarve has the potential to become an even more valuable tool for surgeons, educators, illustrators and students.

\section*{Code Availability}

The full source code for AnatomyCarve is available as a Unity3D package: \url{https://github.com/andrey-titov/AnatomyCarve}. 

A limited-functionality re-implementation as a 3D Slicer module is available at: \url{https://github.com/andrey-titov/SlicerAnatomyCarve}.

\section*{Acknowledgments}
This work was funded by the Natural Sciences and Engineering Research Council of Canada, Discovery program (RGPIN-2020-05084), as well as by the Fonds de recherche du Québec – Nature et technologies (B2X Doctoral Scholarship), Application 334501 (https://doi.org/10.69777/334501).

%\section*{Code Availability}

%The code of AnatomyCarve is publicly available at our GitHub repository \url{https://github.com/andrey-titov/AnatomyCarve}.

%\section*{Supplemental Materials}

%A video demonstrating the functionalities of AnatomyCarve is provided along with the article.

% \section*{Figure Credits}
% Fig. \ref{fig:main_a} and \ref{fig:main_b} left panels adapted from  J. Maclise, Surgical Anatomy \cite{maclise_surgical_2008}, which is in the public domain.

\bibliographystyle{IEEEtran}

\bibliography{references}

\newpage
 
\vspace{11pt}

\begin{IEEEbiography}[{\includegraphics[width=1in,height=1.25in,clip,keepaspectratio]{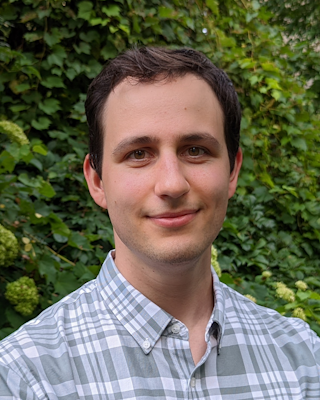}}]{Andrey Titov} received his BCompSc and MCompSc degrees in computer science at Concordia University (Montreal) in 2018 and 2020. In 2021, he started working towards his Ph.D. degree in computer science at \'Ecole de Technologie Sup\'erieure (Montreal). He is a member of the Multimedia Lab at ÉTS and the Applied Perception Lab at Concordia University. His research interests include medical visualization, human-computer interaction, volumetric rendering and virtual reality.
\end{IEEEbiography}

\vspace{11pt}

\begin{IEEEbiography}[{\includegraphics[width=1in,height=1.25in,clip,keepaspectratio]{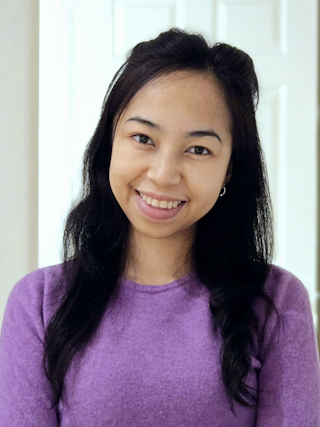}}]{Tina N. H. Nantenaina} received the M.D. degree from the Faculty of Medicine, University of Antananarivo (Madagascar), in 2020. In 2024, she obtained the MScA degree in Health Technology from École de technologie supérieure (Montréal) and the Engineering degree in e‑health from the École d’ingénieur en informatique de la santé (France). Currently, she is a Research Assistant in the Neuro‑iX Lab at ÉTS and part of the research group there. Her research interests include surgical planning and simulation using virtual reality.
\end{IEEEbiography}

\vspace{11pt}

\begin{IEEEbiography}[{\includegraphics[width=1in,height=1.25in,clip,keepaspectratio]{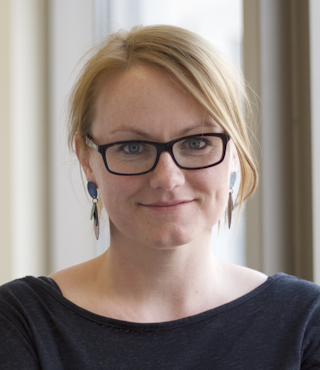}}]{Marta Kersten-Oertel} received the BSc (Honours) degree in Computer Science and the BA degree in Art History from Queen’s University (Kingston) in 2002. In 2005 she completed her MSc in Computer Science at Queen’s University. After working as a research assistant at the GRaphisch-Interaktive Systeme at the University of T\"ubingen (Germany), in 2015 she received the PhD degree in Biomedical Engineering at McGill University (Montreal). She is an Associate Professor in Computer Science and Software Engineering, Concordia University Research Chair in Applied Perception and are FRQ Research Scholar. Her research is focused on developing and evaluating new visualization techniques, and display and interaction methods specifically for health and clinical applications.
\end{IEEEbiography}

\vspace{11pt}

\begin{IEEEbiography}[{\includegraphics[width=1in,height=1.25in,clip,keepaspectratio]{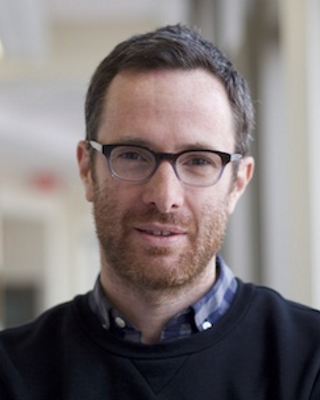}}]{Simon Drouin} received the Bachelor's degree in Computer Engineering from École Polytechnique de Montréal, in 2000, the MSc degree in computer science from McGill University, in 2008, where he focused on applications of computer vision. In 2016, he received the PhD degree in Biomedical Engineering from McGill University (Montreal). Between different study periods, he worked as a software developer at Ubisoft Entertainment inc., Janro Imaging Labs inc., the National Film Board of Canada (NFB) and the Montreal Neurological Institute. Currently, he is a Professor at the Software and Information Technology Engineering Department at \'Ecole de Technologie Sup\'erieure and part the Multimedia Lab there. His research focuses on human-computer interaction involving real-time graphics rendering.
\end{IEEEbiography}

\vfill

\end{document}